\documentclass[a4paper, amsfonts, amssymb, amsmath, reprint, showkeys, nofootinbib, twoside, onecolumn,notitlepage]{revtex4-1}

\def\mnras{{MNRAS}}             
\def\prc{{Phys.~Rev.~C}}        
\def\prd{{Phys.~Rev.~D}}        
\usepackage{amsmath,amstext}
\usepackage[T1]{fontenc}
\usepackage{amssymb}
\usepackage{graphicx}
\usepackage{ae,aecompl}

\usepackage{hyperref}
\usepackage{amsmath}
\usepackage{amssymb}
\usepackage{mathtools}
\usepackage{bm}
\usepackage{cleveref}
\usepackage{tensor}
\usepackage{braket}
\usepackage{enumitem}
\usepackage{mhchem}
\usepackage{amsthm}

\usepackage{color}

\newcommand{\spc}{\quad \quad \quad}

\def\be{\begin{equation}}
\def\ee{\end{equation}}
\def\beq{\begin{eqnarray}}
\def\eeq{\end{eqnarray}}

\theoremstyle{definition}

\theoremstyle{theorem}

\begin{document}
\title{Stability and causality of Carter's multifluid theory}
\author{L.~Gavassino}
\affiliation{Nicolaus Copernicus Astronomical Center, Polish Academy of Sciences, ul. Bartycka 18, 00-716 Warsaw, Poland}

\begin{abstract}
Stability and causality are studied for linear perturbations about equilibrium in Carter's multifluid theory. Our stability analysis is grounded on the requirement that the entropy of the multifluid, plus that of the environment, must be maximised at equilibrium. This allows us to compute a quadratic Lyapunov functional, whose positive definiteness implies stability. Furthermore, we verify explicitly that, also for multifluids, thermodynamic stability implies linear causality. As a notable stability condition, we find that the entrainment matrix must always be positive definite, confirming a widespread intuition.
\end{abstract}

\maketitle

\section{Introduction}

Carter's multifluid theory \cite{carter1991,cool1995,Carter_starting_point} is the hydrodynamic framework currently adopted for modelling superfluid-normal mixtures in full general relativity \cite{Prix_single_vortex,andersson2007review,Termo}. It extends the notion of perfect fluid to interacting systems in which non-diffusive relative flows can survive over hydrodynamic time-scales. As such, it finds application in neutron star physics \cite{langlois98,andersson_comer2000,GavassinoIordanskii2021}: dense hadronic matter is believed to be a mixture of several chemical components, some of which are superfluid (free to spin at different rates \cite{sourie_glitch2017,antonelli+2018,Geo2020}). Furthermore, the ability to describe non-diffusive out-of-equilibrium fluxes makes Carter's theory well suited for modelling dissipation beyond Fick's law \cite{noto_rel,Lopez09,Lopez11}, elevating the formalism to a pillar of Relativistic Extended Irreversible Thermodynamics \cite{Jou_Extended,BulkGavassino,
GavassinoUEIT2021,CamelioGavassino2022}. 

Despite the relevance of Carter's multifluid approach, both for non-equilibrium statistical mechanics and for astrophysical modelling, very little is known about its mathematical properties. In particular, to date no systematic study of its stability and causality properties has ever been carried out. In other words, we do not know under which conditions the initial value formulation of Carter's theory is reliable, and produces physically meaningful solutions. Only few specific hydrodynamic models, built using Carter's approach, have been shown to be reliable (or non-reliable \cite{OlsonRegularCarter1990}). This has always been done by invoking some mathematical correspondence \cite{PriouCOMPAR1991} with the Israel-Stewart theory \cite{Israel_Stewart_1979}, whose stability-causality properties are well known \cite{Hishcock1983}. Unfortunately, such correspondence is limited to theories with only two currents (entropy and particles), and cannot be extended to, e.g., neutron star hydrodynamics (which requires at least three currents: entropy, protons, and neutrons).

This is a serious issue, not only because we do not know if the currently adopted multifluid models are reliable, but also because we have no idea of which factors contribute to make a theory stable, and which approximations, instead, may be harmful. The case of Carter's ``regular theory'' for heat conduction is emblematic: Carter correctly identified the origin of the instability of the theories of \citet{Eckart40} and \citet{landau6}; he formulated a new theory with the precise goal of fixing such pathologies \cite{noto_rel}; nevertheless, the resulting theory turned out to be still unstable (for some realistic equations of state \cite{OlsonRegularCarter1990}). 

To make the problem even more dramatic, there is the fact that determining the ``multifluid equation of state'' from microscopic calculations becomes more and more difficult as the number of relevant currents increases. In fact, if a theory possesses $N$ currents, the number of off-diagonal ``entrainment coefficients'' (i.e. non-viscous couplings \cite{prix2004}) that one needs to compute is $N(N-1)/2$.
Considering that the computation of a single entrainment coefficient is already a challenge, the evaluation all 6 coefficients in a model for neutron star cores (that has 4 currents: entropy, superconducting protons, superfluid neutrons, and normal electrons) is out of the question. Hence, one is forced to rely on approximations, e.g. by setting some entrainment coefficients to zero \cite{GavassinoKhalatnikov2021}. But we do not know, in general, if an approximation of this kind is permitted, or if it spoils the reliability of the whole theory.

The goal of this paper is to finally derive the stability and causality conditions of Carter's multifluid theory. In order to do it, we will invoke some recent developments in the relativistic theory of hydrodynamic stability. 
In particular, we will make use of the following facts: 
\begin{itemize}
\item The stability conditions of a theory which is mathematically consistent with the second law of thermodynamics can be derived from the requirement that the entropy is maximised at equilibrium (Gibbs stability criterion \cite{GavassinoGibbs2021}). Conversely, a fluid theory which predicts that the equilibrium state is a saddle point of the entropy is necessarily unstable (if the second law is obeyed \cite{GavassinoLyapunov_2020}).
\item A theory that respects the Gibbs stability criterion, and is consistent with the second law, is linearly causal \cite{GavassinoCausality2021}.
\item A dissipative field theory, which is stable in one reference frame, is causal if and only if it is stable in all reference frames \cite{GavassinoSuperluminal2021}, see also \cite{BemficaDNDefinitivo2020}.
\end{itemize}

We adopt the signature $(-,+,+,+)$ and work in natural units $c=k_B=1$. Unless otherwise specified, we adopt Einstein's summation convention for spacetime indices ($a,b,c,d$), chemical indices ($X,Y$), and charge indices ($I,\lambda,\gamma$).

\section{Setting the stage}

In this section, we provide a quick overview of Carter's multifluid theory using the generating function approach~\cite{GavassinoKhalatnikov2021}. Then, we apply the Gibbs stability criterion \cite{GavassinoGibbs2021} in the ``$\text{fluid}+\text{environment}$'' formulation  \cite{GavassinoCausality2021}.

\subsection{Carter's theory}\label{carterstheory}

A multifluid is a mixture of several distinct chemical species, which are free to flow independently, and in different directions. In the general case discussed here, each chemical species contributes with 4 degrees of freedom (1 for its density, 3 for its flow). In the present paper, we will work in the ``pressure $\&$ momenta'' representation \cite{lebedev1982}, while the most common formulation of Carter's theory is in the ``master function $\&$ currents'' representation \cite{andersson2007review}. The two formulations are equivalent: they are connected by a change of variables \cite{Carter_starting_point,Prix_single_vortex}.

We introduce a chemical index $X$, which runs over all the relevant chemical species of the system\footnote{The notion of ``chemical species'' here is very general: it may include particle species (like protons and neutrons), but also quasi-particle species (like phonons and rotons), or even macroscopic effective currents with no particle-like microscopic counterpart (like the entropy).}, including the entropy, given by $X=s$.
The state of a multifluid, in Carter's theory, can be completely characterised by a collection of covector fields $\mu^X_a$ (one for each species $X$), called ``momenta'' of the multifluid. The equation of state is given in terms of a generalised pressure $P$, written as a function of the momenta and of the metric:
\begin{equation}
P=P(\mu^X_a,g^{ab}) \, .
\end{equation}
The central postulate of the theory is that the fluxes of the multifluid are determined by the following differential \cite{GavassinoKhalatnikov2021}:
\begin{equation}\label{dP}
\dfrac{d (\sqrt{-g} \, P)}{\sqrt{-g}} = -n_X^a \, d\mu^X_a - \dfrac{1}{2} T_{ab} \, dg^{ab} \, ,
\end{equation}
where $n_X^a$ is the four-current of the chemical species $X$, and $T^{ab}$ is the (total) stress-energy tensor of the multifluid. Einstein's summation convention is adopted for repeated chemical indices. 

Equation \eqref{dP} can be used to write $T^{ab}$ in terms of $\mu^X_a$, $n_X^a$, and $P$. In fact, Lorentz-covariance demands that $P$ is a function of the scalars $g^{ab}\mu^X_a \mu^Y_b$. Therefore, there must be a symmetric matrix $\mathcal{K}_{XY}$ such that
\begin{equation}\label{dP2}
dP= -\dfrac{1}{2} \, \mathcal{K}_{XY} \, d(g^{ab}\mu^X_a \mu^Y_b) \, .
\end{equation}
Recalling that \cite{MTW_book}
\begin{equation}
\dfrac{d\sqrt{-g}}{\sqrt{-g}} = \dfrac{g^{ab} \, dg_{ab}}{2} = -\dfrac{g_{ab} \, dg^{ab}}{2} \, ,
\end{equation}
we can combine \eqref{dP} with \eqref{dP2}, obtaining (note that the symmetry of $\mathcal{K}_{XY}$ implies the symmetry of $T_{ab}$)
\begin{equation}\label{Idue}
\begin{split}
& n_X^a = \mathcal{K}_{XY} \mu^{Ya} \\
& T_{ab} = P g_{ab} + \mathcal{K}_{XY} \mu^{X}_a \mu^Y_b \, . \\
\end{split}
\end{equation}
Combining these two equations, we obtain the well-known formula
\begin{equation}\label{Tabcanonico}
T\indices{^a _b} = P g\indices{^a _b} + n_X^a \mu^X_b \, ,
\end{equation}
which justifies the interpretation of $P$ as a pressure, and of $\mu^X_a$ as ``momenta''. Often, it is convenient to rewrite the first equation of \eqref{Idue} in the form
\begin{equation}\label{entrainment}
\mu^X_a = \mathcal{K}^{XY}n_{Ya} \, ,
\end{equation}
where $\mathcal{K}^{XY}$ (the so called ``entrainment matrix'') is defined as the matrix inverse of $\mathcal{K}_{XY}$:
\begin{equation}
\mathcal{K}^{XY} \mathcal{K}_{YZ} = \delta\indices{^X _Z},
\end{equation}
and is in turn symmetric.

Note that, if the currents are all collinear with each other, i.e. $n_X^a=n_X u^a$ ($u^a$ is the collective four-velocity: $u^a u_a =-1$), then also the momenta are collinear, i.e. $\mu^X_a = \mu^X u_a$, with $\mu^X = \mathcal{K}^{XY}n_Y$, and the stress-energy tensor takes the standard perfect-fluid form, with energy density $\rho=n_X \mu^X-P$ (Euler relation). Furthermore, restriction of the differential \eqref{dP} to states of this kind (taking $dg^{ab}=0$) leads to the Gibbs-Duhem equation: $dP=n_X d\mu^X$ (use the fact that $u_a du^a=0$, which follows from the normalization of $u^a$). Therefore, a multifluid in which all the species flow together is a multi-constituent perfect fluid \cite{Termo}.

\subsection{Allowed processes}\label{ALLUCE}

Consider an isolated system, comprised of a multifluid in weak contact with an environment H (``heat bath''), which evolves hydrodynamically from a state 1 to a state 2. Then, we know that [notation: $\Delta A=A_2-A_1$]
\begin{equation}\label{transition}
\Delta S+\Delta S^H \geq 0  \spc \Delta Q_I + \Delta Q_I^H =0 \, ,
\end{equation}
where quantities with label H refer to the environment, while the others refer to the multifluid. The first condition is the second law of thermodynamics ($S$ is the entropy), while the second is the charge conservation: $Q_I$ are all the conserved charges of the system, such as the baryon number and the four-momentum. An ideal heat bath is defined as an effectively infinite system, whose equation of state can be approximated as \cite{GavassinoTermometri} (we adopt Einstein's convention also for the label $I$) 
\begin{equation}\label{bath}
S^H= -\alpha^I_\star Q^H_I +\text{const} \, ,
\end{equation}
where $\alpha^I_\star$ are some constants: they are the \textit{fixed} intensive properties of the bath. Combining \eqref{transition} with \eqref{bath}, we obtain 
\begin{equation}\label{deltaallowed}
\Delta (S+\alpha^I_\star Q_I) \geq 0 \, .
\end{equation}
In conclusion, a hydrodynamic process is allowed only if it does not involve a decrease in the function $\Phi = S+\alpha^I_\star Q_I$. Note that, here, $S$ and $Q_I$ refer to the multifluid alone (while the constants $\alpha^I_\star$ characterise the environment). Hence, we can use multifluid hydrodynamics to estimate $\Phi$. In particular, $\Phi$ can be expressed as the integral
\begin{equation}\label{phiInSigma}
\Phi[\Sigma] = \int_\Sigma \phi^a \, d\Sigma_a \, ,
\end{equation}
where $\Sigma$ is an arbitrary space-like 3D-surface covering the whole mutifluid, $d\Sigma_a$ is the volume one-form (with standard orientation \cite{MTW_book}: $d\Sigma_0 >0$), and 
\begin{equation}\label{phiat}
\phi^a = s^a +\alpha^I_\star \, J\indices{^a_I} \, ,
\end{equation}
with $s^a=n_s^a$ (the entropy current), and $J\indices{^a_I}$ are the conserved currents (whose charges are $Q_I$).

A multifluid has two types of conserved currents, for which we will use respectively the indices $\lambda$ and $\gamma$, hence $I=\{ \lambda, \gamma \}$. First, there are the currents that can be expressed as 
\begin{equation}
J\indices{^a _\lambda} = q\indices{^X _\lambda} \, n_X^a \, ,
\end{equation}
where $q\indices{^X _\lambda}$ are some constant coefficients, representing the amount of conserved charge $\lambda$ carried by the species $X$. Currents of this kind are, for example, the baryon current, the lepton current, and the electric current. Clearly, the entropy does not carry any conserved charge, hence $q\indices{^s _\lambda }=0$. Secondly, there are the currents of the form \cite{Hawking1973}
\begin{equation}
J\indices{^a _\gamma} = K\indices{^b _\gamma} \, T\indices{^a _b} \, ,
\end{equation}
where $K\indices{^b _\gamma}$ are the Killing vector fields of the spacetime (treated here as a fixed background). For example, if the spacetime is Minkowski, there is a conserved current for each of the 10 generators of the Poincar\'e group, whereas, if the spacetime is Kerr, there are only 2 such currents: energy and angular momentum current. Thus, equation \eqref{phiat} can be rewritten as (Einstein's convention applies also to $\lambda$ and $\gamma$)
\begin{equation}\label{phiaaat}
\phi^a = s^a +\alpha^\lambda_\star \, q\indices{^X _\lambda} \, n_X^a + \alpha^\gamma_\star \, K\indices{^b _\gamma} \, T\indices{^a _b} \, .
\end{equation}
Introducing the compact notation
\begin{equation}\label{alphastarxxx}
\alpha^X_\star = \delta\indices{^X _s} + \alpha^\lambda_\star \, q\indices{^X _\lambda} \spc \beta^b_\star = \alpha^\gamma_\star \, K\indices{^b _\gamma} \, ,
\end{equation}
equation \eqref{phiaaat} becomes
\begin{equation}\label{barbossa153}
\phi^a = \alpha^X_\star n_X^a + \beta^b_\star \, T\indices{^a _b} \, .
\end{equation}
In the following, we will assume that the Killing vector field $\beta^b_\star$ (which is the so called ``inverse-temperature vector'' \cite{Becattini2016}) is time-like future-directed, so that we can express it as
\begin{equation}\label{uT}
\beta^b_\star = \dfrac{u^b}{T} \spc \text{with} \quad u^b u_b=-1 \, ; \quad u^0,T>0 \, .
\end{equation} 
As we shall see, the unit vector field $u^b$ can be interpreted as the (local) equilibrium conglomerate flow velocity of the multifluid, while the scalar field $T$ is the (local) equilibrium temperature\footnote{Note, however, that $\beta^b_\star$, as defined in \eqref{alphastarxxx}, exists (and is the same) also far from equilibrium. In fact, $\alpha_\star^\gamma$ are constant properties of the bath, which do not depend on the state of the multifluid, while $K\indices{^b _\gamma}$ are fixed by the choice of metric. Hence, we should think of $u^b$ and $T$ as externally-imposed fixed parameters, which acquire a hydrodynamic meaning only at equilibrium. The same is true for $\alpha_\star^X$.}. 

Note that, if some currents are superfluid, we can define some quasi-conserved topological charges (i.e. winding numbers \cite{Termo}), whose existence is responsible for the long life of the superflow \cite{AndreevMelnikovski2004,GavassinoKhalatnikov2021}. However, in the present paper, we will assume that such quasi-conservation laws are eventually broken at the length/time-scales of interest (and we will not includes such charges among the ``$Q_I$''). This assumption is justified whenever a large number of vortices can be generated and can travel across the multifluid, inducing a vortex-mediated mutual friction \cite{langlois98,GavassinoIordanskii2021}, which effaces all relative flows, at equilibrium.

\subsection{Thermodynamic equilibrium}\label{Thermodynamicequilibrium}

Since the quantity $\Phi$ cannot decrease, the state of thermodynamic equilibrium is the state that maximises $\Phi$, for a given background metric $g^{ab}$ and a given environment, which play the role of external conditions. Hence, from now on, we will work at fixed $g^{ab}$ and $\alpha_\star^I$. As a consequence, also $\alpha_\star^X$ and $\beta_\star^b$ are fixed ($K\indices{^b _\gamma}$ are determined by $g^{ab}$). 

Let us consider a smooth one-parameter family, $ \mu^X_a(\epsilon)$, of solutions of the fluid equations, for which $\epsilon=0$ is the equilibrium state (for the given $g^{ab}$ and $\alpha_\star^I$). For each value of the parameter $\epsilon$ (and for each choice of 3D-surface $\Sigma$), we can compute the quantity $\Phi(\epsilon)$, which can be differentiated with respect to $\epsilon$ [notation: $\dot{A}=dA/d\epsilon$]. Then, the maximality of $\Phi$ at equilibrium implies 
\begin{equation}\label{bigphidot}
\dot{\Phi}(0)=0   \, ,
\end{equation} 
for any choice of one-parameter family of solutions (defined as above), and for any space-like 3D-surface $\Sigma$, covering the whole multifluid. Recalling equations \eqref{phiInSigma}, \eqref{barbossa153}, \eqref{Tabcanonico}, and \eqref{dP}, the quantities $\Phi$, $\dot{\Phi}$, and $\ddot{\Phi}$ can be respectively expressed as flux-integrals of the following currents (for any value of $\epsilon$):
\begin{equation}\label{letrederivate}
\begin{split}
& \phi^a = (\alpha^X_\star + \beta^b_\star \mu^X_b) n_X^a + P  \beta^a_\star   \\
& \dot{\phi}^a = (\alpha^X_\star + \beta^b_\star \mu^X_b) \dot{n}_X^a + (\beta^b_\star  n_X^a -n_X^b   \beta^a_\star) \dot{\mu}^X_b \\
& \ddot{\phi}^a = (\alpha^X_\star + \beta^b_\star \mu^X_b) \ddot{n}_X^a + (\beta^b_\star  n_X^a -n_X^b   \beta^a_\star) \ddot{\mu}^X_b +2\beta^b_\star \dot{\mu}^X_b \dot{n}_X^a -  \beta^a_\star \dot{n}_X^b  \dot{\mu}^X_b \, .  \\
\end{split}
\end{equation}
Since equation \eqref{bigphidot} must be respected for any choice of one-parameter family, and for any choice of $\Sigma$, we must require $\dot{\phi}^a(0) =0$, which implies that the equilibrium state ($\epsilon=0$) satisfies the conditions
\begin{equation}\label{eofmklmbgokf}
  -\beta^b_\star \mu^X_b = \alpha^X_\star  \spc \beta^{[b}_\star  n_X^{a]}=0 \, .
\end{equation}
Recalling equation \eqref{uT}, we can rewrite the equilibrium conditions above as follows:
\begin{equation}\label{ubmub}
\dfrac{-u^b \mu^X_b}{T} =  \alpha^X_\star \spc n_X^b = n_X u^b \, .
\end{equation}
The second condition has a simple interpretation: at global thermodynamic equilibrium, all chemical components flow with the same conglomerate four-velocity $u^b$. Furthermore,  since $\beta_\star^b=u^b/T$ is a Killing vector field (hence, $\nabla^a \beta_{\star}^b+\nabla^b \beta_{\star}^a=0$), the equilibrium velocity $u^b$ satisfies the condition
\begin{equation}\label{nablaua}
\nabla_{(a}u_{b)}= u_{(b}\nabla_{a)} \ln T \, ,
\end{equation}
which can be projected orthogonally to $u^b$ (using the projection tensor $h^{ab}=g^{ab}+u^a u^b$), giving
\begin{equation}
h^{ca}h^{db}\nabla_{(a}u_{b)} =0 \, .
\end{equation}
This means that the equilibrium conglomerate fluid motion is shear-less and expansion-less, as one would expect.

Let us analyse the first condition of \eqref{ubmub}. First, we note that $\alpha_\star^s=1$ [see equation \eqref{alphastarxxx}, and recall that $q\indices{^s _\lambda}=0$], so that $T_b := \mu^s_b$ (which is often called ``thermal covector'') satisfies the equation
\begin{equation}\label{emfvlce}
T=-u^b  T_b \, .
\end{equation} 
Furthermore, for $X \neq s$, the first equation of \eqref{ubmub} reduces to the well-known equilibrium condition $\mu^X/T=\text{const}$ ($\mu^X$ is the chemical potential of the species $X$), provided that we make the identification
\begin{equation}\label{erfomkler}
\mu^X =-u^b \mu^X_b \, .
\end{equation} 
Equation \eqref{erfomkler} is more than a formal identification: it is a rigorous thermodynamic identity. In fact, at equilibrium the multifluid is a perfect fluid (see subsection \ref{carterstheory}), with collective flow velocity $u^a$, rest-frame densities $n_X$, and chemical potentials $-u^b \mu^X_b$.


We can make an additional observation. Assume that, in the multifluid, there is a possibility for the following chemical reaction to occur:
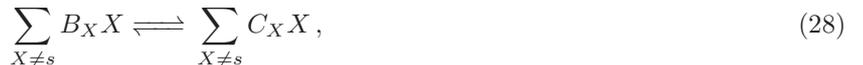
\begin{equation}
\sum_{X\neq s} B_X X \ce{ <=> }  \sum_{X\neq s} C_X X \, , 
\end{equation}
where $B_X$ and $C_X$ are some stoichiometric coefficients. Clearly, the reaction is forbidden if it does not conserve all the charges $Q_\lambda$. Hence, any allowed reaction needs to satisfy the charge-balance conditions
\begin{equation}
\sum_{X\neq s} B_X \, q\indices{^X _\lambda}  = \sum_{X\neq s} C_X \, q\indices{^X _\lambda}  \spc \forall \, \lambda \, .
\end{equation}
Contracting this equation with $\alpha_\star^\lambda$, recalling \eqref{alphastarxxx}, \eqref{ubmub}, and \eqref{erfomkler}, we obtain the usual chemical equilibrium condition
\begin{equation}
\sum_{X\neq s} B_X \mu^X = \sum_{X\neq s} C_X \mu^X \, .
\end{equation}
It follows that the state of thermodynamic equilibrium, computed by maximising the function $\Phi$, is also a state of chemical equilibrium, with respect to \textit{any} reaction which is compatible with the conservation laws.

As a final remark, note that at equilibrium ($\epsilon=0$) equations \eqref{phiInSigma}, \eqref{letrederivate}, and \eqref{eofmklmbgokf} can be combined to give
\begin{equation}
\Phi(0) = \int_\Sigma P  \beta^a_\star \, d\Sigma_a  \, ,
\end{equation}
which is equation (3.17) of \citet{GibbonsHawking1977}\footnote{Note that \citet{GibbonsHawking1977} adopt the non-standard orientation for the volume 1-form: $d\Sigma^0 >0$.}. This implies that the thermodynamic properties of the equilibrium state are consistent with the predictions of quantum statistical mechanics.

\subsection{Stability criterion}\label{criteriuz}

In subsection \ref{Thermodynamicequilibrium}, we have identified the equilibrium state by demanding that it makes $\Phi$ stationary: $\dot{\Phi}(\epsilon=0)=0$. However, we still need to make sure that $\epsilon=0$ is a genuine maximum of $\Phi$. In other words, we need to show that, for any $\epsilon$, the functional [notation: $\delta A = A(\epsilon)-A(0)$]
\begin{equation}\label{Emenodeltaphi}
E = -\delta \Phi  
\end{equation}
is non-negative definite, and vanishes only at equilibrium. Recalling equation \eqref{phiInSigma}, we can express $E$ as an integral:
\begin{equation}\label{edsigma}
E[\Sigma]=\int_\Sigma E^a \, d\Sigma_a \, , \, \quad \text{with } \,  E^a=-\delta \phi^a.
\end{equation} 
Close to $\epsilon=0$ (i.e. close to equilibrium) we can expand $E^a$ to second order [recall that $\dot{\phi}^a(0)=0$]:
\begin{equation}
E^a =\phi^a(0)-\phi^a(\epsilon)= -\dfrac{1}{2} \ddot{\phi}^a(0) \, \epsilon^2 + \mathcal{O}(\epsilon^3) \, .
\end{equation}
Using the third equation of \eqref{letrederivate}, we can write\footnote{The contributions to $\ddot{\phi}^a(0)$ proportional to $\ddot{n}_X^a$ and $\ddot{\mu}^X_b$ vanish, because of equation \eqref{eofmklmbgokf}. Thus, $E^a$ is quadratic in ``$\dot{A}$ quantities'', and we can make the replacements $ \dot{A}(0) \, \epsilon + \mathcal{O}(\epsilon^2) = A(\epsilon)-A(0)=\delta A$, because the corresponding error to $E^a$ is of order $\epsilon^3$.}
\begin{equation}\label{proptioTE}
TE^a = \dfrac{u^a}{2} \delta n_X^b \delta \mu^X_b - u^b \delta n_X^a \delta \mu^X_b + \mathcal{O}(\epsilon^3) \, .
\end{equation}
This is the same current $E^a$ that we obtained in \cite{GavassinoGibbs2021}. The goal of this paper is to study under which conditions the current $E^a$ gives rise to a positive definite functional $E$, for any perturbation $\delta \mu^X_b$. Under such conditions, $E$ plays the role of a square-integral norm of the perturbation. On the other hand, $E=\Phi(0)-\Phi(\epsilon)$, where $\Phi(0)$ is the equilibrium value of $\Phi$, which is a constant, while $\Phi(\epsilon)$ depends of the 3D-surface $\Sigma$ upon which it is calculated; in particular, $\Phi(\epsilon)$ is a non-decreasing function of time [see equation \eqref{deltaallowed}]. It follows that the positive-definite norm $E$ can only decrease with time (or stay constant), meaning that small perturbations away from equilibrium cannot grow: the equilibrium state is Lyapunov-stable (for small perturbations). This is true for \textit{any} process which is consistent with the conservation laws and with the second law of thermodynamics: the details of the field equations governing the system are irrelevant (we did not even specify the field equations!), provided that \eqref{transition} holds.

\section{Stability analysis}

In this section, we derive the conditions under which $E$ is positive for all non-vanishing small perturbations. If these conditions are respected, the theory is linearly stable.

\subsection{Equilibrium-frame decomposition}

First of all, it is useful to rewrite \eqref{proptioTE} in a more transparent form. Let us consider a non-equilibrium state $\epsilon \neq 0$. For such state, the density $n_X(\epsilon)$ and the chemical potential $\mu^X(\epsilon)$ are not uniquely defined, because there is no collective flow velocity. However, we can use the \textit{equilibrium} flow velocity $u^a$ [see equation \eqref{uT}], which does not depend on $\epsilon$, to define the non-equilibrium density and chemical potential as follows:
\begin{equation}\label{kmunghiz}
n_X(\epsilon)=-n_X^a(\epsilon) \, u_a  \spc \mu^X(\epsilon)=- \mu^X_a(\epsilon) \,  u^a \, .
\end{equation}
Their interpretation is simple: they are the density and chemical potential of the species $X$ in the non-equilibrium state $\epsilon$, as measured in the \textit{equilibrium} local rest-frame of the multifluid (defined by $u^a$). Note that $n_X(0)$ coincides with $n_X$ [see equation \eqref{ubmub}], while $\mu^X(0)$ coincides with $\mu^X$ [see equation \eqref{erfomkler}].

Next, we can use $u^a$ to decompose the non-equilibrium currents and momenta as follows [recall equation \eqref{entrainment}]:
\begin{equation}\label{nexepsilon}
\begin{split}
& n_X^a(\epsilon)= n_X(\epsilon) \, u^a + j_X^a(\epsilon) \\
& \mu^X_a(\epsilon)= \mu^X(\epsilon) \, u_a + \mathcal{K}^{XY}(\epsilon) \, j_{Ya}(\epsilon) \, , \\
\end{split}
\end{equation}
where
\begin{equation}
j_X^a(\epsilon) \, u_a =0 \, .
\end{equation}
According to this decomposition, $j_X^a(\epsilon)$ is the non-equilibrium flux of the species $X$, measured in the equilibrium rest frame. Clearly, $j_X^a(0)=0$ [see equation \eqref{ubmub}]. Now, we can use equation \eqref{nexepsilon} to rewrite the variations $\delta n_X^a$ and $\delta \mu^X_a$ as follows [recall our notation: $A(\epsilon)-A(0)= \delta A$]:
\begin{equation}\label{ddnexepsilon}
\begin{split}
& \delta n_X^a=  u^a \, \delta n_X + \delta j_X^a \\
& \delta \mu^X_a= u_a \,  \delta \mu^X  + \mathcal{K}^{XY} \delta j_{Ya} + \mathcal{O}(\epsilon^2) \, , \\
\end{split}
\end{equation} 
where, in the second equation, $\mathcal{K}^{XY}$ is evaluated at equilibrium\footnote{To first-order, one has $\delta (\mathcal{K}^{XY}j_{Ya})=\mathcal{K}^{XY} \delta j_{Ya}  +j_{Ya} \delta \mathcal{K}^{XY}$, but the second term vanishes, because $j_{Ya}=0$ at equilibrium.} (i.e., at $\epsilon=0$).

We can make one last observation. Clearly, the variation $\delta \mu^X$ can be expressed (to first order) as
$
\delta \mu^X = \rho\indices{^X ^Y _b} \, \delta n_Y^b 
$,
for some background matrix $\rho\indices{^X ^Y _b}$. However, at equilibrium all hydrodynamic vectors are collinear to $u^a$, so that the only expression for $\rho\indices{^X ^Y _b}$ which is compatible with the symmetries of the system is $\rho\indices{^X ^Y _b} = -\rho^{XY} u_b$, for some matrix $\rho^{XY}$. Therefore, recalling equation \eqref{ddnexepsilon}, we obtain
\begin{equation}\label{muxxx}
\delta \mu^X = \rho^{XY} \delta n_Y + \mathcal{O}(\epsilon^2) \, .
\end{equation}
On the other hand, when all the perturbed currents are collinear to $u^a$, the perturbed multifluid is indistinguishable from a multiconstituent perfect fluid. We can invoke this correspondence to conclude that
\begin{equation}
\rho^{XY}= \dfrac{\partial^2 \rho}{\partial n_X \partial n_Y} \, ,
\end{equation}
where the derivative is computed in the perfect-fluid limit (i.e. with the constraint $n_X^a =n_X u^a$), and $\rho= n_X \mu^X -P$ is the prefect-fluid energy density. Plugging \eqref{ddnexepsilon} and \eqref{muxxx} into \eqref{proptioTE}, we finally arrive at
\begin{equation}\label{Lafinale}
TE^a = \dfrac{u^a}{2} \bigg[ \rho^{XY} \delta n_X \delta n_Y + \mathcal{K}^{XY} \delta j_X^b \delta j_{Yb} \bigg] + \rho^{XY} \delta j_X^a \, \delta n_Y \, ,
\end{equation}
where we are neglecting third-order terms in $\epsilon$.

%
%

\subsection{Stability in the rest frame}\label{restuz}

Let us assume that the equilibrium state is non-rotating. Then, we can foliate the space-time with space-like 3D-surfaces $\Sigma$, which are everywhere orthogonal to the equilibrium flow velocity $u^a$. For such foliation, the volume one-form $d\Sigma_a$ in \eqref{edsigma} can be rewritten as (recall that we are adopting the standard orientation: $d\Sigma_0>0$)
\begin{equation}
d\Sigma_a = -u_a \, dV \, ,
\end{equation}
where $dV>0$ is the metric volume element of $\Sigma$. Then, the requirement that $E$ should be positive definite reduces to the condition (we multiply by $T>0$ for convenience) 
\begin{equation}\label{eTea}
e := -T E^a u_a >0 \, , 
\end{equation}
on any spacetime point where the perturbation does not vanish. Plugging \eqref{Lafinale} into \eqref{eTea}, we obtain
\begin{equation}
2e =  \rho^{XY} \delta n_X \delta n_Y + \mathcal{K}^{XY} \delta j_X^b \delta j_{Yb} \, .
\end{equation}
Clearly, $2e$ is positive for any choice of perturbation $\{ \delta n_X , \delta j_X^b \}$ if and only if $\rho^{XY}$ and $\mathcal{K}^{XY}$ are positive definite (symmetric) matrices. In fact, $\rho^{XY} \delta n_X \delta n_Y$ is a quadratic form in $\delta n_X$, whereas $\mathcal{K}^{XY} \delta j_X^b \delta j_{Yb}$ (working in a local Lorentz frame comoving with $u^a$) is a sum of three independent quadratic forms, respectively in $\delta j_X^1$, $\delta j_X^2$ and $\delta j_X^3$.

\subsection{Stability in a generic reference frame}

If the equilibrium state is rotating, it is impossible to find a 3D-surface $\Sigma$ which is everywhere orthogonal to $u^a$. In this case, we are forced to work with an arbitrary space-like 3D-surface, for which we have
\begin{equation}
d\Sigma_a = -\tilde{u}_a \, dV \, ,
\end{equation}
where $\tilde{u}^a$ is the future-directed (time-like) unit normal to $\Sigma$, and $dV>0$ is the metric volume element of $\Sigma$. This time, the requirement that $E$ should be positive definite reduces to the condition
\begin{equation}\label{tigrisdellegallie}
e = T \dfrac{E^a \tilde{u}_a}{u^b \tilde{u}_b}  >0 \, ,
\end{equation}
on any spacetime point where the perturbation is non-vanishing, and for any choice of $\tilde{u}^a$. This is equivalent to saying that $E^a$ is time-like future-directed for any non-vanishing perturbation. Let us decompose the four-vector $\tilde{u}^a$ in the equilibrium rest-frame (defined by $u^a$),
\begin{equation}\label{decuppo}
\tilde{u}^a = -u^b\tilde{u}_b (u^a+w^a) \, ,
\end{equation}
with $w^a u_a=0$ ($w^a$ is the three-velocity, relative to $u^a$, of observers moving along $\tilde{u}^a$), and define the projection tensor
\begin{equation}\label{gammupo}
\gamma_{ab} = g_{ab}+u_a u_b - \dfrac{w_a w_b}{w^2} \, , 
\end{equation}
where $w^2=w^a w_a \in [0,1)$, because $\tilde{u}^a$ is time-like. Plugging \eqref{Lafinale} and \eqref{decuppo} into \eqref{tigrisdellegallie}, and using \eqref{gammupo}, we obtain
\begin{equation}
2e = \rho^{XY}(\delta n_X-w_a \delta j_X^a)(\delta n_Y-w_b \delta j_Y^b) + \mathcal{K}^{XY} \gamma_{ab} \delta j_X^a \delta j_Y^b +  \big( \mathcal{K}^{XY}-w^2 \rho^{XY} \big)\dfrac{w_a \delta j_X^a \, w_b \delta j^b_Y}{w^2} \, .
\end{equation}
We note that the quantities
\begin{equation}
\{ \, \delta n_X-w_a \delta j_X^a \, , \, \gamma_{ab}\delta j_X^b \, , \, w_a \delta j_X^a \, \}  
\end{equation}
are independent from each other, and constitute a parameterization of the degrees of freedom of the perturbation. Hence, $2e$ is positive for all non-vanishing perturbations if and only if $\rho^{XY}$, $\mathcal{K}^{XY}$ and $\mathcal{K}^{XY}-w^2 \rho^{XY}$ are positive definite matrices. This must be true for any space-like 3D-surface, and, therefore, for any $w^2 \in [0,1)$. Since $\rho^{XY}$ is positive definite, it is evident that the matrices $\mathcal{K}^{XY}-w^2 \rho^{XY}$ are all positive definite provided that $\mathcal{K}^{XY}-\rho^{XY}$ is positive definite.

In conclusion, a multifluid is stable to linear perturbations if the symmetric matrices
\begin{equation}
\rho^{XY}, \spc \mathcal{K}^{XY}, \spc \mathcal{K}^{XY}-\rho^{XY}
\end{equation}
are positive definite. If these matrices are only non-negative definite, the stability of the theory is uncertain, because the sign of $E$ is determined by higher-order terms (in $\epsilon$). If, on the other hand, any of these matrices fails to be non-negative definite, the theory is unstable, because a perturbation that pushes $E$ below zero cannot evolve back to equilibrium (as $E$ is a non-increasing function of time). 

Let us make some final remarks:
\begin{itemize}
\item The positive definiteness of $\mathcal{K}^{XY}$ follows directly from the positive definiteness of $\rho^{XY}$ and $ \mathcal{K}^{XY}-\rho^{XY}$.  
\item Clearly, also the matrix $\mathcal{K}_{XY}$ (the inverse of $\mathcal{K}^{XY}$) is positive definite.
\item As we said in subsection \ref{criteriuz}, the present criterion for stability is valid for any choice of field equations, provided that $\nabla_a s^a \geq 0$ and $\nabla_a J\indices{^a_I}=0$ [express every ``$\, \Delta A \,$'' in \eqref{transition} using Gauss theorem]. This is a crucial point: no matter how we prescribe the dissipative equations of the multifluid (with or without reactions, resistivities, and gradient-dependent forces \cite{GavassinoRadiazione}), the stability conditions are always the same, provided that the second law of thermodynamics is respected, as a strict mathematical inequality.
\item All our calculations are valid in the absence of viscous stresses, which would enter the constitutive relations as corrections to \eqref{Tabcanonico}. However, the functional $E$ of a viscous theory is that of the inviscid theory, plus a piece that vanishes for vanishing perturbations to the viscous stresses. It follows that, if $E$ fails to be positive definite in the inviscid limit, the same is true for the full viscous theory. We can conclude that the present stability conditions are also \textit{necessary} (but not sufficient) stability conditions for all viscous multifluids.
\item From the previous two points, we can draw a useful lesson: dissipative effects such as chemical reactions and resistivities \cite{carter1991}, which result only in a modification of the field equations (without altering the constitutive relations), cannot affect the stability properties of the system \cite{GavassinoGibbs2021,GavassinoUEIT2021}. On the other hand, dissipative effects such as viscosity, which modify the constitutive relations, can induce instabilities (if modelled incorrectly).
\end{itemize}

\section{Some quick applications}\label{quickona}

In this section, we derive the most interesting stability conditions of some selected multifluid models. We do not perform the whole stability analysis directly, because it is straightforward (albeit tedious). Instead, our aim is to develop an intuition of what causes some theories to be unstable, and which strategies we can adopt to fix the instabilities.

\subsection{Perfect fluid at zero chemical potential}\label{lesempietto}

Let us consider a ``multifluid'' whose only current is the entropy current $s^a$. Such multifluid is simply a finite-temperature perfect fluid with zero chemical potential, and is often used as a minimal model for the quark-gluon plasma \cite{FlorkowskiReview2018}. The matrices $\rho^{XY}$ and $\mathcal{K}^{XY}$ have only one dimension:
\begin{equation}
\rho^{ss}= T/c_v \spc \mathcal{K}^{ss} = T/s \, ,
\end{equation}
where $c_v = T \, ds/dT$ is the heat capacity per unit volume. The rest frame stability conditions are simply $c_v >0$ and $s>0$, which are well-established thermodynamic inequalities. The last condition is
\begin{equation}
\mathcal{K}^{ss}-\rho^{ss}= \dfrac{T}{sc_v} (c_v-s)>0 \, ,
\end{equation}
which may be expressed as
\begin{equation}
c_s^2 = \dfrac{s}{c_v} < 1 \, ,
\end{equation}
where $c_s^2= dP/d\rho$ is the square of the speed of sound. We have just rediscovered a very general result: thermodynamic stability implies causality \cite{GavassinoCausality2021}. We will explore the causality issue in more detail in subsection \ref{causalitypropriobene}.

\subsection{Inviscid models for heat conduction}\label{Heattuz}

\citet{noto_rel} has shown that several \textit{inviscid} models for heat conduction can be reinterpreted as multifluids, with two currents: $s^a$ (entropy current), and $n^a$ (conserved particle current). On the ordered chemical basis $\{ s^a, n^a \}$, the elements of the entrainment matrix are usually denoted by
\begin{equation}
\mathcal{K}^{XY}=
\begin{bmatrix}
   \mathcal{C} & \mathcal{A}  \\
    \mathcal{A} & \mathcal{B}  \\
\end{bmatrix} \, ,
\end{equation} 
where $\mathcal{C}$ stands for ``caloric coefficient'', $\mathcal{B}$ stands for ``bulk coefficient'', and $\mathcal{A}$ stands for ``anomaly coefficient'' \cite{cool1995}. The stress-energy tensor \eqref{Tabcanonico} can be decomposed using $\mathcal{C}$, $\mathcal{B}$, and $\mathcal{A}$ as follows:
\begin{equation}\label{adueadue}
T^{ab}= P g^{ab} + \mathcal{C} \, s^a s^b + \mathcal{A} \, (n^a s^b+s^a n^b) + \mathcal{B} \, n^a n^b \, .
\end{equation}
Different theories for heat conduction postulate a different geometrical structure for the stress-energy tensor and, consequently, adopt different formulas for the entrainment coefficients. 

For example, the Eckart theory \cite{Eckart40} posits that $\mathcal{C}=0$ (no contribution to $T^{ab}$ proportional to $s^a s^b$), and $\mathcal{A} \neq 0$. This implies that $\mathcal{K}^{XY}$ has negative determinant: the Eckart theory is unstable, even in the rest frame. The case of the Landau-Lifshitz theory \cite{landau6} is more interesting. It posits that the stress-energy tensor has a perfect-fluid structure (in the inviscid limit), namely, there is a vector field $v^a$ such that $T^{ab}=P g^{ab}+v^a v^b$. Comparing this constraint with \eqref{adueadue}, we see that $\pm v^a$ must coincide with one of the two vectors $\sqrt{\mathcal{C}} \, s^a \pm \sqrt{\mathcal{B}} \, n^a$, and the determinant
\begin{equation}
\det || \mathcal{K}^{XY} || = \mathcal{C}\mathcal{B}-\mathcal{A}^2
\end{equation}
must vanish. It follows that $\mathcal{K}^{XY}$ has one vanishing eigenvalue. This puts the Landau-Lifshitz theory at the boundary between stable and unstable theories, in the rest frame. On the other hand, since $\rho^{XY}$ is positive definite, the matrix $\mathcal{K}^{XY}- w^2 \rho^{XY}$ will have one negative eigenvalue for any $w^2 \neq 0$. This explains why the Landau-Lifshitz theory is unstable in any reference frame which is non-comoving with the equilibrium four-velocity $u^a$.

To fix the problems of the aforementioned theories, \citet{noto_rel} formulated the ``regular theory'', defined by the condition $\mathcal{A}=0$. For such theory, the entrainment matrix takes a very simple form [plug \eqref{entrainment} into \eqref{kmunghiz}, and set $\mathcal{A}=0$]
\begin{equation}\label{kappuzzo}
\mathcal{K}^{XY}=
\begin{bmatrix}
   T/s & 0  \\
    0 & \mu/n  \\
\end{bmatrix} \, .
\end{equation} 
We note that, for $\mathcal{K}^{XY}$ to be positive definite, we must have $\mu /n >0$. This is not a standard thermodynamic inequality: there may be fluids that violate this postulate. However, considering that $\mu$ is the \textit{relativistic} chemical potential (it contains the ``$ \, mc^2 \,$'' contribution), this is not expected to happen in many astrophysical systems. The real problem is the condition $\mathcal{K}^{ss}-\rho^{ss}>0$, which produces the constraint
\begin{equation}\label{Glradiator}
c_v >s \, .
\end{equation}
This requirement is too strong, and is violated even by the non-degenerate Boltzmann gas\footnote{Note that, in \cite{OlsonRegularCarter1990}, $c_v$ and $s$ are quantities per particle, while here they are quantities per unit volume. Obviously, the inequality is the same: one only needs to divide our equation \eqref{Glradiator} by $n$, to recover equation (54) of \cite{OlsonRegularCarter1990}.} \cite{OlsonRegularCarter1990}. For this reason, Carter's regular theory is unstable for many realistic equations of state. The present analysis is mathematically equivalent to that of \citet{OlsonRegularCarter1990}, but it is more straightforward, because it is grounded on the direct study of the entrainment matrix.

Finally, close to equilibrium, also the inviscid Israel-Stewart theory \cite{Israel_Stewart_1979} can be mapped into a multifluid, with entrainment matrix \cite{PriouCOMPAR1991}
\begin{equation}
\mathcal{K}^{XY}=
\begin{bmatrix}
   \beta_1 T^2 & & \dfrac{T}{n} (1-\beta_1 sT)  \\
     & &  \\
  \, \, \dfrac{T}{n} (1-\beta_1 sT) & & \dfrac{\mu}{n} - \dfrac{sT}{n^2} (1-\beta_1 sT) \, \, \\
\end{bmatrix} \, ,
\end{equation} 
where $\beta_1$ is a second-order transport coefficient of the Israel-Stewart theory (see \cite{Hishcock1983} for the definition). In this case, the determinant of the entrainment matrix is
\begin{equation}
\det || \mathcal{K}^{XY} || = \dfrac{T^2}{n^2} \bigg[ \beta_1 (sT+n\mu) -1 \bigg]\, ,
\end{equation}
so that the condition $\det || \mathcal{K}^{XY} || >0$ implies
\begin{equation}\label{gbomfl}
\beta_1 > \dfrac{1}{\rho +P} \, ,
\end{equation}
where we have invoked the equilibrium identity $sT + n\mu = \rho +P$. The inequality \eqref{gbomfl} is a well-known stability condition of the Israel-Stewart theory. This is not surprising: our current $E^a$, given in equation \eqref{proptioTE}, is the inviscid limit of the current $E^a$ used by Hiscock and Lindblom to assess the stability of the Israel-Stewart theory \cite{Hishcock1983}, and coincides with equation (31) of \citet{OlsonRegularCarter1990}, see \cite{GavassinoGibbs2021} for the proof. Thus, the stability conditions are the same. For example, it is straightforward to show that $\mathcal{K}^{ss}-\rho^{ss}=T^2 \, \Omega_5(1)>0$, see equation (42) of \cite{OlsonRegularCarter1990}.

\subsection{Relativistic two-fluid model for superfluid Helium}\label{4he}

A one-component superfluid, such as $^4$He, can be modelled, in the non-dissipative limit \cite{GavassinoKhalatnikov2021}, as a multifluid with two currents: $s^a$ (entropy current), and $n^a$ (conserved particle current). The resulting theory is the relativistic generalization of Landau's two-fluid model for superfluidity\footnote{This theory has the same currents as a model for heat conduction (see subsection \ref{Heattuz}), but the field equations, whose details are irrelevant for our purposes, are completely different \cite{noto_rel,cool1995}.} \cite{carter92}. On the ordered chemical basis $\{ s^a,n^a \}$, the entrainment matrix $\mathcal{K}^{XY}$ can be expressed in terms of the Landau superfluid and normal ``mass densities'' (respectively $\rho_S$ and $\rho_N$) as follows \cite{cool1995}:
\begin{equation}\label{Tarzan}
\mathcal{K}^{XY}=
\begin{bmatrix}
   \dfrac{\rho_N}{s^2} + \dfrac{\rho_S}{s^2}\bigg(1-\dfrac{n\mu}{\rho_S} \bigg)^2 & & \, \, \dfrac{\mu}{s} \bigg( 1-\dfrac{n\mu }{\rho_S} \bigg)  \\
     & &  \\
   \dfrac{\mu}{s} \bigg( 1-\dfrac{n\mu }{\rho_S} \bigg) & & \dfrac{\mu^2}{\rho_S} \\
\end{bmatrix} \, ,
\end{equation}
which implies
\begin{equation}\label{Tarzan2}
\det || \mathcal{K}^{XY} ||= \dfrac{\mu^2 \rho_N}{s^2 \rho_S} \, ,  \spc \quad sT+ n\mu = \rho_S + \rho_N \, .
\end{equation}
Equations \eqref{Tarzan} and \eqref{Tarzan2} are evaluated in the comoving limit, which is considered to be, in the present paper, the only ``proper'' equilibrium state (see the discussion at the end of subsection \ref{ALLUCE}). Recalling that in the comoving limit $sT+ n\mu =\rho+P$, the second equation of \eqref{Tarzan2} implies $\rho+P = \rho_S + \rho_N$. Therefore, in relativity, $\rho_S$ and $\rho_N$ are not partitions of the rest mass density $mn$: they are partitions of the enthalpy density $\rho+P$. Indeed, it is $\rho+P$ (and not $mn$) that determines the inertia of a relativistic fluid \cite{MTW_book}.

Let us derive some stability conditions. It is easy to show that $\mathcal{K}^{XY}$ is positive definite if and only if
\begin{equation}
\rho_S >0 \, , \spc \rho_N >0 \, ,
\end{equation}
which are well-established thermodynamic inequalities (valid for any superfluid), see equation (16) of \citet{AndreevMelnikovski2004}. Furthermore, if $\mathcal{K}^{XY}-\rho^{XY}$ is positive definite, then $\mathcal{K}^{XX}>\rho^{XX}$ (for any $X$), so that
\begin{equation}\label{ledueloroloro}
\dfrac{\rho_N}{s^2} + \dfrac{\rho_S}{s^2}\bigg(1-\dfrac{n\mu}{\rho_S} \bigg)^2 > \dfrac{T}{c_v} \, , \spc  \dfrac{\mu^2}{\rho_S} > \dfrac{\partial \mu}{\partial n} \bigg|_s \, .
\end{equation}
To understand the physical meaning of these conditions, let us work in the low temperature limit ($\rho_N \ll \rho_S$, and $sT \ll n\mu$), assuming that the elementary excitations are phonons. Then, the second equation of \eqref{Tarzan2} becomes $\rho_S \approx n\mu$, and we have the identities \cite{landau9}
\begin{equation}
s = \dfrac{c_v}{3} = \dfrac{ \rho_N}{T} (c_{s1})^2 \, ,  \spc \text{ with } \quad  (c_{s1} )^2 := \dfrac{n}{\mu}\dfrac{\partial \mu}{\partial n} \bigg|_{s=0} \, . 
\end{equation}
Plugging these approximations into \eqref{ledueloroloro}, we obtain the following inequalities:
\begin{equation}
\dfrac{(c_{s1})^2}{3} <1  \, , \spc (c_{s1})^2<1 \, . 
\end{equation}
On the other hand, $c_{s1}$ is the speed of first sound, while $c_{s2}=c_{s1}/\sqrt{3}$ is the speed of second sound \cite{landau6,cool1995}. Again, thermodynamic stability implies causality.

\subsection{Entrainment in superfluid neutron stars}

A minimal model for a superfluid neutron star builds on three currents: $s^a$ (entropy current), $n_n^a$ (neutron current), and $n_p^a$ (proton current). Hence, the matrices $\mathcal{K}^{XY}$ and $\rho^{XY}$ have dimension 3. On the other hand, if $\mathcal{K}^{XY}$, $\rho^{XY}$, and $\mathcal{K}^{XY}-\rho^{XY}$ are positive definite, also their $2 \times 2$ sub-blocks $\{ n,p \}$ must be positive definite. Here, we will compute some of the stability conditions associated with these sub-blocks.

If the multifluid is in beta equilibrium (namely, $\mu^p = \mu^n =: \mu$), the entrainment matrix, in the ordered basis $\{ n,p \}$, can be written as follows \cite{antonelli+2018}:
\begin{equation}
\mathcal{K}^{XY}=\mu 
\begin{bmatrix}
   \dfrac{1-\varepsilon_n}{n_n} &  & \dfrac{\varepsilon_n }{n_p}  \\
   & & \\
   \dfrac{\varepsilon_p}{n_n}  & & \dfrac{1-\varepsilon_p}{n_p}  \\
\end{bmatrix} \, ,
\end{equation} 
where $n_n \varepsilon_n = n_p \varepsilon_p$, which follows from the symmetry of $\mathcal{K}^{XY}$. Stability in the rest frame demands $\varepsilon_n <1$, $\varepsilon_p <1$, and
\begin{equation}
\det || \mathcal{K}^{XY} || =  \dfrac{\mu^2}{n_n n_p} (1-\varepsilon_n - \varepsilon_p) >0 \, .
\end{equation}
Comparing the conditions above, we find that the most stringent is
\begin{equation}
\varepsilon_n < \dfrac{n_p}{n_p+n_n} \, .
\end{equation}
This same condition (valid both in the core and the crust of neutron stars) was obtained by \citet{chamelhaensel2006} and \citet{carter_macro_2006} in a Newtonian setting, by demanding that the dynamical contribution to the energy density be positive definite, see equation (67) of \cite{chamelhaensel2006} and equation (4.27) of \cite{carter_macro_2006}. As we can see, it remains valid also in relativity.

Finally, the positive-definiteness of $\mathcal{K}^{XY}-\rho^{XY}$ produces two notable inequalities:
\begin{equation}
\dfrac{1}{1-\varepsilon_n} \, \dfrac{\partial \ln \mu^n}{\partial  \ln n_n}\bigg|_{n_p} <1 \spc \dfrac{1}{1-\varepsilon_p} \, \dfrac{\partial \ln \mu^p}{\partial  \ln n_p}\bigg|_{n_n} <1 \, . 
\end{equation}
As we shall see in subsection \ref{causalitypropriobene}, these conditions are necessary to ensure causality.


\subsection{Rau-Wasserman model for superfluid neutron stars}

\citet{RauWasserman2020} have constructed a multifluid model for superfluid neutron stars with 7 currents: $X \in \{s, n,p,\bar{n},\bar{p}, e,m \}$, representing respectively entropy, normal neutrons, normal protons, superfluid neutrons, superconducting protons, electrons, and muons. Since they assume that $\mathcal{A}^{s\bar{n}}=\mathcal{A}^{s\bar{p}}=0$, we believe that, by ``normal'' and ``superfluid'' component, they mean the Landau-type normal and superfluid part of the total currents, which are precisely defined as the choice of chemical basis in which the entrainment with the entropy is zero \cite{carter92}. They also postulate that $\mathcal{A}^{Xe}=\mathcal{A}^{Xm}=0$ for $X \neq s$, and $\mathcal{A}^{\bar{n}\bar{p}}=\mathcal{A}^{n\bar{p}}=\mathcal{A}^{\bar{n} p}=\mathcal{A}^{np}$, so that we have
\begin{equation}
\mathcal{K}^{XY}=
\begin{bmatrix}
\mathcal{C} & \mathcal{A}^{sn} & \mathcal{A}^{sp} & 0 & 0 & \mathcal{A}^{se} & \mathcal{A}^{sm}  \\
\mathcal{A}^{sn} & \mathcal{B}^n & \mathcal{A}^{np} & \mathcal{A}^{n\bar{n}} & \mathcal{A}^{np} & 0 & 0  \\
\mathcal{A}^{sp} & \mathcal{A}^{np} & \mathcal{B}^p & \mathcal{A}^{np} & \mathcal{A}^{p\bar{p}} & 0 & 0  \\
0 & \mathcal{A}^{n\bar{n}} & \mathcal{A}^{np} & \mathcal{B}^{\bar{n}} & \mathcal{A}^{np} & 0 & 0  \\
0 & \mathcal{A}^{np} & \mathcal{A}^{p\bar{p}} & \mathcal{A}^{np} & \mathcal{B}^{\bar{p}} & 0 & 0  \\
\mathcal{A}^{se} & 0 & 0 & 0 & 0 & \mathcal{B}^e & 0  \\
\mathcal{A}^{sm} & 0 & 0 & 0 & 0 & 0 & \mathcal{B}^m  \\
\end{bmatrix} \, .
\end{equation}
Let us see the most straightforward stability conditions. Clearly, all the coefficients $\mathcal{B}^X$ must be positive. Furthermore, we have that
\begin{equation}
\begin{split}
& \mathcal{C}> \max \bigg\{ \dfrac{(\mathcal{A}^{sn})^2}{\mathcal{B}^n} , \dfrac{(\mathcal{A}^{sp})^2}{\mathcal{B}^p}, \dfrac{(\mathcal{A}^{se})^2}{\mathcal{B}^e}, \dfrac{(\mathcal{A}^{sm})^2}{\mathcal{B}^m}  \bigg\}\\
& |\mathcal{A}^{np}| < \min \{ \sqrt{\mathcal{B}^n \mathcal{B}^p} , \sqrt{\mathcal{B}^n \mathcal{B}^{\bar{p}}}, \sqrt{\mathcal{B}^{\bar{n}} \mathcal{B}^p} , \sqrt{\mathcal{B}^{\bar{n}} \mathcal{B}^{\bar{p}}} \} \\
& |\mathcal{A}^{n\bar{n}}| < \sqrt{\mathcal{B}^n \mathcal{B}^{\bar{n}}} \\
& |\mathcal{A}^{p\bar{p}}| < \sqrt{\mathcal{B}^p \mathcal{B}^{\bar{p}}} \, , \\
\end{split}
\end{equation}
to ensure the positive definiteness of the relative $2 \times 2$ blocks. Also, the positive-definiteness of the matrix $\mathcal{K}^{XY}-\rho^{XY}$ produces the notable conditions
\begin{equation}\label{inguo}
\begin{split}
& \mathcal{C}> T/c_v \\
& \mathcal{B}^X > \partial \mu^X /\partial n_X \, . \\
\end{split}
\end{equation}
Finally, we can make the following observations:
\begin{itemize}
\item Rau and Wasserman insist on the necessity of keeping \textit{all} the entrainment couplings between the entropy and the normal currents different from zero (i.e. $\mathcal{A}^{sX}\neq 0$ for $X\neq\bar{p},\bar{n}$), to ensure stability and causality. This conjecture is motivated by the analogy with Carter's regular theory, in which the removal of the entrainment with the entropy causes the failure of the stability condition $\mathcal{K}^{ss}-\rho^{ss}>0$. However, inspection of equation \eqref{kappuzzo} reveals that the real problem of the regular theory is not ``$\, \mathcal{A}^{sn}=0 \,$'' itself, but $\mathcal{C}=T/s$, which leads to the unphysical condition $c_v >s$ [see \eqref{inguo}]. But, to remove the constraint $\mathcal{C}=T/s$, we need only one of the coefficients $\mathcal{A}^{sX}$ to be non-vanishing, not necessarily all of them. Hence, the model may be simplified further.
\item Rau and Wasserman suggest that a further simplification could be to set $\mathcal{B}^n=\mathcal{B}^{\bar{n}}=\mathcal{A}^{n\bar{n}}$. However, they add that this may lead to a contradiction with the condition of chemical equilibrium: $\mu^n = \mu^{\bar{n}}$ \cite{carter92}. Actually, postulating $\mathcal{B}^n=\mathcal{B}^{\bar{n}}=\mathcal{A}^{n\bar{n}}$ would lead to even bigger problems. In fact, it would imply that $\mathcal{B}^n\mathcal{B}^{\bar{n}}-(\mathcal{A}^{n\bar{n}})^2=0$. Thus, the sub-block $\{ n, \bar{n}\}$ of $\mathcal{K}^{XY}$ would fail to be positive definite, and the theory would suffer from the same instabilities that plague the Landau-Lifshitz theory for heat conduction. 
\item We remark that the present stability analysis is valid only in the inviscid limit. On the other hand, the Rau-Wasserman model \cite{RauWasserman2020} has the ambition of including also viscous corrections. In equation (127), they postulate viscous stresses which are linear in the spatial gradients, as in the Landau-Lifshitz theory for viscosity. This makes the viscous model acausal\footnote{Before equation (127), \citet{RauWasserman2020} claim that their viscous model is causal, invoking a mathematical correspondence with \cite{PriouCOMPAR1991}. However, equations (161) and (162) of \cite{PriouCOMPAR1991} present the standard relaxation-time terms (proportional to $\beta_0$ and $\beta_2$), which have been neglected in \cite{RauWasserman2020}. This approximation makes the Rau-Wasserman viscous model acausal.} and, therefore, unstable \cite{GavassinoSuperluminal2021}. 
\end{itemize}

\section{Physical interpretation}

In this section, we explore in more detail the physical origin of the stability conditions discussed in the present paper. To simplify the discussion, we will assume that the equilibrium state is homogeneous, namely $\nabla_a u^b = \nabla_a n_X =0$.

\subsection{Theory of fluctuations}

Because of the interaction with H, the equilibrium density operator $\hat{\sigma}_{\text{eq}}$ of the multifluid is grand-canonical \cite{GibbonsHawking1977}:
\begin{equation}
\hat{\sigma}_{\text{eq}}= \dfrac{\exp(\alpha_\star^I \hat{Q}_I)}{Z} \spc \text{with } \, Z= \text{Tr} \exp(\alpha_\star^I \hat{Q}_I) \, .
\end{equation}
Here, $\hat{Q}_I$ are the quantum operators associated with the conserved charges $Q_I$. Each macrostate of the multifluid has an associated projector $\hat{\mathcal{P}}$, which projects on the Hilbert subspace defined by all the microscopic realizations of the macrostate. The entropy $S$ of the macrostate is given by Boltzmann's formula: $\exp (S)= \text{Tr} \, \hat{\mathcal{P}}= \text{``number of microscopic realizations''}$. Hence, the probability of observing the multifluid in a given macrostate is
\begin{equation}
\mathcal{P} = \text{Tr}(\hat{\sigma}_{\text{eq}} \hat{\mathcal{P}}) \approx \dfrac{\exp(\alpha_\star^I Q_I)}{Z} \, \text{Tr} \, \hat{\mathcal{P}} = \dfrac{\exp(S+\alpha_\star^I Q_I)}{Z} = \dfrac{\exp (\Phi)}{Z} \, ,
\end{equation}
where $Q_I$ is the macroscopic value of the charge in the given macrostate ($\hat{Q}_I \hat{\mathcal{P}} \approx Q_I \hat{\mathcal{P}}$). Recalling equation \eqref{Emenodeltaphi}, we can conclude that the grand-canonical probability distribution of the thermodynamic fluctuations is \cite{GavassinoCausality2021}
\begin{equation}
\mathcal{P} \propto \exp(-E) \, .
\end{equation}
Restricting our attention to homogeneous configurations (as measured in the equilibrium rest frame), and choosing the 3D-surface $\Sigma$ to be orthogonal to $u^a$, we obtain
\begin{equation}\label{0jomo0mo}
\mathcal{P} \propto \exp \int_\Sigma E^a u_a dV = \exp \bigg[ -\dfrac{V\rho^{XY}}{2T} \delta n_X \delta n_Y -\dfrac{V \mathcal{K}^{XY}}{2T} \delta j_X^b \delta j_{Yb}  \bigg] \, ,
\end{equation}
where $V=\int_\Sigma dV$ is the total volume of the multifluid. 
As we can see, the matrices $\mathcal{\rho}^{XY}$ and $\mathcal{K}^{XY}$ determine the typical size of the statistical fluctuations of respectively $ n_X$ and $ j_X^b$. Their positive definiteness is necessary, to guarantee that the equilibrium macrostate is the most probable macrostate. Furthermore, equation \eqref{0jomo0mo} tells us that the matrix $\mathcal{K}_{XY}$ (the inverse of the entrainment matrix) has a simple statistical interpretation:
\begin{equation}
\braket{\delta j_X^a \delta j_Y^b} = \dfrac{Th^{ab}}{V} \, \mathcal{K}_{XY}  \, .
\end{equation}
This is the generalization of ``$\, \braket{\delta v^1 \delta v^1}=T/M \,$''  \cite{landau5} to relativistic multifluids.

\subsection{Acoustic properties of the multifluid}\label{causalitypropriobene}

Using the theory of fluctuations, we have explained the positive definiteness of $\rho^{XY}$ and $\mathcal{K}^{XY}$. We are left with the task of explaining the positive definiteness of $\mathcal{K}^{XY}-\rho^{XY}$. Two observations come to our aid:
\begin{itemize}
\item As we anticipated in the introduction, a dissipative theory that is stable in one reference frame is causal if and only if it is stable in all reference frames \cite{GavassinoSuperluminal2021}. On the other hand, the positive definiteness of $\rho^{XY}$ and $\mathcal{K}^{XY}$ guarantees the stability in the rest frame. Hence, demanding that $\mathcal{K}^{XY}-\rho^{XY}$ is positive definite must be equivalent to demanding causality.
\item In \cite{GavassinoCausality2021}, we have shown that, if $E^a$ is time-like future-directed (and the second law is valid), localised perturbations cannot exit the future lightcone. This also suggests that the positive definiteness of $\mathcal{K}^{XY}-\rho^{XY}$ implies causality.
\end{itemize} 
We can conclude that the matrix $\mathcal{K}^{XY}-\rho^{XY}$ determines the acoustic (i.e. causal) properties of the multifluid. This has already been verified explicitly in the case of a theory with a single current (see subsection \ref{lesempietto}), and for superfluid Helium at low temperature (see subsection \ref{4he}). Now we will prove it in full generality. 

In General Relativity, ``causality'' means that any change of initial data in a region of space $\mathcal{R}$ can never exit $J^+(\mathcal{R})$, which is the causal future (or domain of influence) of $\mathcal{R}$ \cite{Wald,Hawking1973,BemficaCausality2018}. In practice, this amounts to requiring that the characteristics of the field equations do not exit the lightcone \cite{Susskind1969,Hishcock1983,BemficaDNDefinitivo2020}. 
In Carter's theory, the dissipative field equations are usually postulated to be (no summation over $X$)
\begin{equation}
2n_X^a \nabla_{[a} \mu^X_{b]} + \mu^X_b \nabla_a n_X^a =\mathcal{R}^X_b \, ,
\end{equation}
where $\mathcal{R}^X_b$ are some dissipative hydrodynamic forces. It is common practice to assume that $\mathcal{R}^X_b$ do not depend on the gradients \cite{GavassinoUEIT2021,GavassinoKhalatnikov2021}. Hence, if our goal is to determine the characteristic speeds of the multifluid, we can just work in the non-dissipative limit\footnote{The characteristic speeds depend only on the principal part of field equations. Thus, the forces $\mathcal{R}^X_b$ do not affect the characteristic determinant (and the causal properties) of the multifluid.} ($\mathcal{R}^X_b=0$), so that, for linear deviations from equilibrium, the field equations can be decomposed as follows:
\begin{equation}\label{menatana2}
\nabla_a \delta n_X^a=0  \spc u^a (\nabla_a \delta \mu^X_b -\nabla_b \delta \mu^X_a)=0 \, .
\end{equation}
Invoking the decompositions \eqref{ddnexepsilon} and \eqref{muxxx}, we obtain [for homogeneous backgrounds]
\begin{equation}\label{mentana}
\begin{split}
& u^a \nabla_a \delta n_X +h^{ab}\nabla_a \delta j_{Xb} =0\\
& \mathcal{K}^{XY} u^a \nabla_a \delta j_{Yb} + \rho^{XY} h\indices{^a _b} \nabla_a \delta n_Y =0 \, , \\
\end{split}
\end{equation}
where $h^{ab}=g^{ab}+u^a u^b$. The equations above can be combined together, giving 
\begin{equation}\label{gargantua}
(\mathcal{K}^{XY}u^a u^b - \rho^{XY} h^{ab})\nabla_a \nabla_b \delta n_Y =0 \, .
\end{equation}
We search for plane-wave solutions of the form
\begin{equation}\label{soundwave}
\delta n_Y(x^a)= \delta n_Y(0) \, \exp(ik_a x^a) \, .
\end{equation}
We can impose $k_a \in \mathbb{R}$, because equations \eqref{menatana2} are non-dissipative. Hence, \eqref{soundwave} models a sound wave, and we are allowed to write [we just express ``$\, \omega^2 =c_s^2 k^2 \,$'' using covariant language]
\begin{equation}\label{pantagruel}
(u^a k_a)^2 = c_s^2 \, h^{ab}k_a k_b \, .
\end{equation}
Plugging \eqref{soundwave} into \eqref{gargantua}, and invoking \eqref{pantagruel}, we finally obtain
\begin{equation}\label{aragorn}
 \big( \rho^{XY} - c_s^2 \mathcal{K}^{XY} \big)\delta n_Y =0 \, .
\end{equation}
Thus, the speeds of sound squared are the generalised eigenvalues of $\rho^{XY}$ with respect to $\mathcal{K}^{XY}$. This has two consequences. The first consequence is that, if $\mathcal{K}^{XY}-\rho^{XY}$ is positive definite, all such eigenvalues are smaller than one, namely $c_s^2<1$. The second consequence is that the theory is dispersion-free: the eigenvalues $c_s^2$ do not depend on the value of $h^{ab}k_a k_b$. On the other hand, it is well-known that, in dispersion-free theories, $c_s$ are also the characteristic speeds of the system \cite{Krotscheck1978,FoxKuper1970,Rauch_book,Pu2010}. Therefore, we have that $\text{``characteristic speed''}=c_s<1$, provided that $\mathcal{K}^{XY}-\rho^{XY}$ is positive definite. This is precisely what we wanted to prove.

In Appendix \ref{Intotheory}, we study the sound-wave solutions of \eqref{mentana} in more detail, for the interested reader.

\subsection{The stability conditions are intuitive!}

In this final subsection, we aim to convince the reader that the stability conditions discussed here could be ``guessed'' without making any explicit calculation. This article has just provided a rigorous systematization of very intuitive ideas, which were already scattered throughout the literature.

First of all, note that the inequality \eqref{deltaallowed} is just the general-relativistic analogue of $\Delta(U-T_\text{eq} S +P_\text{eq}V) \leq 0 $ \cite{Stuekelberg1962,Israel_2009_inbook}, which was used by \citet{landau5} to derive the standard ``textbook inequalities'' (e.g. positivity of specific heats and compressibilities) that are common to all extensive thermodynamic systems. This means that all such ``textbook inequalities''  must be obeyed also by multifluids, and they are a direct consequence of the positive definiteness of $\rho^{XY}$. Interested readers can see \citet{Hishcock1983}, Section III.c, for a detailed list of these universal inequalities [equations (71)-(101)], and for a graphical representation of those equations of state that are consistent with them (in FIG.1). Unfortunately, the analysis of \citet{Hishcock1983} assumes only one independent chemical constituent, which corresponds to a multifluid with two currents: $n^a$ and $s^a$. When there are more currents, one needs to complement the aforementioned inequalities with the conditions for diffusive and chemical stability discussed in section 12.4 of 
\citet{PrigoginebookModernThermodynamics2014}, whose relativistic generalization is discussed in \cite{GavassinoGibbs2021,CamelioGavassino2022}. 

Secondly, one can easily realise that the stability conditions arising from $\rho^{XY}$ only refer to perturbations for which all chemical components still flow along the equilibrium four-velocity $u^a$. On the other hand, in hydrodynamics, we need also to make sure that the fluid is stable against spontaneous acceleration\footnote{Spontaneous accelerations are a well-known pathology of Eckart's theory \cite{Eckart40}: the fluid likes to increases its flow velocity in one direction by ``pushing'' a lot of heat in the opposite direction to conserve the linear momentum \cite{GavassinoLyapunov_2020}.}. From a thermodynamic perspective, this corresponds to requiring that the ``effective kinetic energy''
$
K :=\frac{1}{2}\mathcal{K}^{XY} \delta j_X^b \delta j_{Yb}
$
is positive definite \cite{chamelhaensel2006}, so that the spontaneous generation of relative flows costs free energy, and it is, therefore, entropically disfavoured [see equation \eqref{0jomo0mo}]. Indeed, this simple idea was already clear to Carter, who suggested the interpretation of $\mathcal{K}^{XY}$ as the ``inertia matrix'' of the multifluid (see Section 3 of \citet{carter92}, and in particular FIG.1).

The last ingredient is causality, which is a necessary and sufficient condition for making ``stability'' a Lorentz-invariant property of a dissipative system \cite{GavassinoSuperluminal2021}. In a multifluid, ``causality'' corresponds to requiring that all the (ultraviolet \cite{CamelioGavassino2022}) speeds of sound are subluminal. On the other hand, the total number of speeds of sound coincides with the total number of independent four-currents (for example, superfluids and heat conducting fluids have both a first and a second sound \cite{rezzolla_book,cool1995,Lopez11}). Thus, for a multifluid with $N$ currents, there are $N$ additional stability conditions, which correspond to demanding the positive definiteness of $\mathcal{K}^{XY}-\rho^{XY}$. The reader can see section 2.4 of \citet{noto_rel} for an example of a causality analysis for a two-component multifluid: the result is consistent with our equation \eqref{aragorn}.

\section{Conclusions}

We have performed the linear stability analysis of Carter's multifluids. We have found that, in order for the equilibrium state to be stable against perturbations, three conditions have to be met:
\begin{enumerate}
\item The Hessian matrix $\rho^{XY}$ of the function $\rho=\rho(n_X)$ (energy density written as function of the densities $n_X$) needs to be positive definite. This condition is a standard thermodynamic requirement, valid for all fluids, and follows directly from the minimum energy principle \cite{Callen_book}.
\item The entrainment matrix $\mathcal{K}^{XY}$ needs to be positive definite. This corresponds to saying that the ``inertia'' of all the components of the multifluid is positive. One may interpret it as the straightforward generalization of the perfect-fluid stability condition $\rho+P>0$ \cite{GavassinoCausality2021}. Indeed, in the perfect-fluid limit, we have $\rho +P = \mathcal{K}^{XY} n_X n_Y$, so that $\rho+P>0$ follows directly from the positive definiteness of the entrainment matrix.
\item The matrix $\mathcal{K}^{XY}-\rho^{XY}$ needs to be positive definite. This stability condition produces mixed inequalities, which relate the entrainment coefficients with standard thermodynamic derivatives (like specific heats and compressibilities). One of these inequalities is violated in Carter's regular theory, originating the instability.
\end{enumerate}
Furthermore, we have shown that the characteristic velocities $c_s$ of the multifluid are solutions of the equation 
\begin{equation}\label{gabibbo}
\det || c_s^2 \, \mathcal{K}^{XY} - \rho^{XY} ||=0 \, .
\end{equation}
If all the stability conditions are respected, $c_s^2 \, \mathcal{K}^{XY} - \rho^{XY}$ has strictly positive determinant for $c_s^2 >1$. Thus, stability implies causality. All the present results are valid also for the theory of \citet{Son2001} and \citet{Gusakov2007} (in the inviscid limit), due to the mathematical correspondence with Carter's theory \cite{Termo}.

\section*{Acknowledgements}

This work was supported by the Polish National Science Centre grant OPUS 2019/33/B/ST9/00942. The author thanks M. Antonelli for reading the manuscript and providing useful comments. 

\appendix

\section{Sound-waves in multifluids}\label{Intotheory}

In this appendix, we show how to compute the sound-wave solutions of a generic multifluid, in the non-dissipative limit (i.e. for $\mathcal{R}^X_b=0$).

\subsection{A preliminary result}

We begin with a useful observation. 
In subsection \ref{restuz}, we saw that $\rho^{XY}$ and $\mathcal{K}^{XY}$ are both symmetric and positive definite. Hence, there are an invertible matrix $\mathcal{N}\indices{^X _A}$ and a positive definite diagonal matrix $\Lambda^{AB}=\lambda^{A} \delta^{AB}$ such that 
\begin{equation}\label{aprity}
\rho^{XY}= \Lambda^{AB} \mathcal{N}\indices{^X _A}\mathcal{N}\indices{^Y _B} \spc \mathcal{K}^{XY}= \delta^{AB} \mathcal{N}\indices{^X _A}\mathcal{N}\indices{^Y _B} \, ,
\end{equation}
where $\delta^{AB}$ is the Kronecker delta-symbol (Einstein's convention for indices $A$ and $B$). What is the physical meaning of the eigenvalues $\lambda^A$? Consider the function
$
f(\lambda)= \det || \, \rho^{XY}- \lambda \mathcal{K}^{XY} || 
$.
Clearly, $f(\lambda)=0$ if and only if $\lambda$ is one of the speeds of sound given in \eqref{aragorn}. On the other hand, using equation \eqref{aprity}, we can rewrite $f(\lambda)$ as
\begin{equation}
f(\lambda) =  \det || \mathcal{K}^{XY}  || \times  \prod_A (\lambda^A-\lambda) \, ,
\end{equation}
which vanishes precisely when $\lambda = \lambda^A$. Hence, the eigenvalues $\lambda^A$ coincide with the squares of the speeds of sound, and we can rewrite them as $\lambda^A=(c_s^A)^2$. 

\subsection{Sound-wave solutions}

Let us plug \eqref{aprity} into \eqref{mentana}. Introducing the notation
\begin{equation}\label{knudsen}
\delta \tilde{n}_A= \mathcal{N}\indices{^X _A} \delta n_X  \spc \delta \tilde{j}^a_A= \mathcal{N}\indices{^X _A} \delta j_X^a \, , 
\end{equation}
the system \eqref{mentana} takes the ``diagonal'' form (no summation over $A$)
\begin{equation}\label{abba}
\begin{split}
& u^a \nabla_a \delta \tilde{n}_A +h^{ab} \nabla_a \delta \tilde{j}_{Ab} =0\\
& u^a \nabla_a \delta \tilde{j}_{Ab} + (c_s^A)^2 h\indices{^a _b} \nabla_a \delta \tilde{n}_A =0 \, . \\
\end{split}
\end{equation}
As we can see, each couple $\{ \delta \tilde{n}_A , \delta \tilde{j}_{Ab} \}$ evolves independently, so that an ``elementary sound-wave'' may be constructed as a plane-wave solution with only one non-vanishing couple $\{ \delta \tilde{n}_A , \delta \tilde{j}_{Ab} \}$.  Such solution can be expressed as
\begin{equation}\label{babuabua}
k_a = k \, (c_s^A u_a + e_a)  \spc  \delta \tilde{j}_{Ab}  = \delta \tilde{n}_A \, c_s^A \, e_b \, , 
\end{equation}
where $k^2=h^{ab} k_a k_b$, and $e^a$ is a normalised ($e^a e_a =1$) space-like vector normal to $u^a$ ($e_a u^a=0$). To compute the perturbations $\{ \delta n_X , \delta j_{Xb} \}$, one can invert equation \eqref{knudsen}.

\subsection{Information current of a sound-wave}

If we plug \eqref{aprity} into \eqref{Lafinale}, and use \eqref{knudsen}, we obtain
\begin{equation}\label{sisplitta}
TE^a = \sum_A \bigg[ \dfrac{u^a}{2}  \lambda^A (\delta \tilde{n}_A)^2 + \dfrac{u^a}{2} \delta \tilde{j}_A^b \delta \tilde{j}_{Ab}  + \lambda^A \delta \tilde{j}_A^a \delta \tilde{n}_A \bigg] \, .
\end{equation}
Evaluating this formula on an elementary plane-wave solution, given by \eqref{babuabua}, we obtain (no summation over $A$)
\begin{equation}
TE^a = (c_s^A \delta \tilde{n}_A)^2 (u^a + c_s^A \, e^a) \, .
\end{equation}
Therefore, $E^a$ points in the direction of propagation of the elementary sound-wave, i.e. $E^a \propto u^a + c_s^A e^a$, in agreement with the interpretation of $E^a$ as the flow of information, transported by the perturbation \cite{GavassinoCausality2021}.

\bibliography{Biblio}

\begin{thebibliography}{61}%
\makeatletter
\providecommand \@ifxundefined [1]{%
 \@ifx{#1\undefined}
}%
\providecommand \@ifnum [1]{%
 \ifnum #1\expandafter \@firstoftwo
 \else \expandafter \@secondoftwo
 \fi
}%
\providecommand \@ifx [1]{%
 \ifx #1\expandafter \@firstoftwo
 \else \expandafter \@secondoftwo
 \fi
}%
\providecommand \natexlab [1]{#1}%
\providecommand \enquote  [1]{``#1''}%
\providecommand \bibnamefont  [1]{#1}%
\providecommand \bibfnamefont [1]{#1}%
\providecommand \citenamefont [1]{#1}%
\providecommand \href@noop [0]{\@secondoftwo}%
\providecommand \href [0]{\begingroup \@sanitize@url \@href}%
\providecommand \@href[1]{\@@startlink{#1}\@@href}%
\providecommand \@@href[1]{\endgroup#1\@@endlink}%
\providecommand \@sanitize@url [0]{\catcode `\\12\catcode `\$12\catcode
  `\&12\catcode `\#12\catcode `\^12\catcode `\_12\catcode `\%12\relax}%
\providecommand \@@startlink[1]{}%
\providecommand \@@endlink[0]{}%
\providecommand \url  [0]{\begingroup\@sanitize@url \@url }%
\providecommand \@url [1]{\endgroup\@href {#1}{\urlprefix }}%
\providecommand \urlprefix  [0]{URL }%
\providecommand \Eprint [0]{\href }%
\providecommand \doibase [0]{http://dx.doi.org/}%
\providecommand \selectlanguage [0]{\@gobble}%
\providecommand \bibinfo  [0]{\@secondoftwo}%
\providecommand \bibfield  [0]{\@secondoftwo}%
\providecommand \translation [1]{[#1]}%
\providecommand \BibitemOpen [0]{}%
\providecommand \bibitemStop [0]{}%
\providecommand \bibitemNoStop [0]{.\EOS\space}%
\providecommand \EOS [0]{\spacefactor3000\relax}%
\providecommand \BibitemShut  [1]{\csname bibitem#1\endcsname}%
\let\auto@bib@innerbib\@empty
\bibitem [{\citenamefont {{Carter}}(1991)}]{carter1991}%
  \BibitemOpen
  \bibfield  {author} {\bibinfo {author} {\bibfnamefont {B.}~\bibnamefont
  {{Carter}}},\ }\href {\doibase 10.1098/rspa.1991.0034} {\bibfield  {journal}
  {\bibinfo  {journal} {Proceedings of the Royal Society of London Series A}\
  }\textbf {\bibinfo {volume} {433}},\ \bibinfo {pages} {45} (\bibinfo {year}
  {1991})}\BibitemShut {NoStop}%
\bibitem [{\citenamefont {Carter}\ and\ \citenamefont
  {Langlois}(1995)}]{cool1995}%
  \BibitemOpen
  \bibfield  {author} {\bibinfo {author} {\bibfnamefont {B.}~\bibnamefont
  {Carter}}\ and\ \bibinfo {author} {\bibfnamefont {D.}~\bibnamefont
  {Langlois}},\ }\href {\doibase 10.1103/PhysRevD.51.5855} {\bibfield
  {journal} {\bibinfo  {journal} {\prd}\ }\textbf {\bibinfo {volume} {51}},\
  \bibinfo {pages} {5855} (\bibinfo {year} {1995})},\ \Eprint
  {http://arxiv.org/abs/hep-th/9507058} {hep-th/9507058} \BibitemShut {NoStop}%
\bibitem [{\citenamefont {{Carter}}\ and\ \citenamefont
  {{Khalatnikov}}(1992)}]{Carter_starting_point}%
  \BibitemOpen
  \bibfield  {author} {\bibinfo {author} {\bibfnamefont {B.}~\bibnamefont
  {{Carter}}}\ and\ \bibinfo {author} {\bibfnamefont {I.~M.}\ \bibnamefont
  {{Khalatnikov}}},\ }\href {\doibase 10.1103/PhysRevD.45.4536} {\bibfield
  {journal} {\bibinfo  {journal} {\prd}\ }\textbf {\bibinfo {volume} {45}},\
  \bibinfo {pages} {4536} (\bibinfo {year} {1992})}\BibitemShut {NoStop}%
\bibitem [{\citenamefont {{Prix}}(2000)}]{Prix_single_vortex}%
  \BibitemOpen
  \bibfield  {author} {\bibinfo {author} {\bibfnamefont {R.}~\bibnamefont
  {{Prix}}},\ }\href {\doibase 10.1103/PhysRevD.62.103005} {\bibfield
  {journal} {\bibinfo  {journal} {\prd}\ }\textbf {\bibinfo {volume} {62}},\
  \bibinfo {eid} {103005} (\bibinfo {year} {2000})},\ \Eprint
  {http://arxiv.org/abs/gr-qc/0004076} {gr-qc/0004076} \BibitemShut {NoStop}%
\bibitem [{\citenamefont {{Andersson}}\ and\ \citenamefont
  {{Comer}}(2007)}]{andersson2007review}%
  \BibitemOpen
  \bibfield  {author} {\bibinfo {author} {\bibfnamefont {N.}~\bibnamefont
  {{Andersson}}}\ and\ \bibinfo {author} {\bibfnamefont {G.~L.}\ \bibnamefont
  {{Comer}}},\ }\href {\doibase 10.12942/lrr-2007-1} {\bibfield  {journal}
  {\bibinfo  {journal} {Living Reviews in Relativity}\ }\textbf {\bibinfo
  {volume} {10}},\ \bibinfo {eid} {1} (\bibinfo {year} {2007})},\ \Eprint
  {http://arxiv.org/abs/gr-qc/0605010} {gr-qc/0605010} \BibitemShut {NoStop}%
\bibitem [{\citenamefont {{Gavassino}}\ and\ \citenamefont
  {{Antonelli}}(2020)}]{Termo}%
  \BibitemOpen
  \bibfield  {author} {\bibinfo {author} {\bibfnamefont {L.}~\bibnamefont
  {{Gavassino}}}\ and\ \bibinfo {author} {\bibfnamefont {M.}~\bibnamefont
  {{Antonelli}}},\ }\href {\doibase 10.1088/1361-6382/ab5f23} {\bibfield
  {journal} {\bibinfo  {journal} {Classical and Quantum Gravity}\ }\textbf
  {\bibinfo {volume} {37}},\ \bibinfo {eid} {025014} (\bibinfo {year}
  {2020})},\ \Eprint {http://arxiv.org/abs/1906.03140} {arXiv:1906.03140
  [gr-qc]} \BibitemShut {NoStop}%
\bibitem [{\citenamefont {{Langlois}}\ \emph {et~al.}(1998)\citenamefont
  {{Langlois}}, \citenamefont {{Sedrakian}},\ and\ \citenamefont
  {Carter}}]{langlois98}%
  \BibitemOpen
  \bibfield  {author} {\bibinfo {author} {\bibfnamefont {D.}~\bibnamefont
  {{Langlois}}}, \bibinfo {author} {\bibfnamefont {D.~M.}\ \bibnamefont
  {{Sedrakian}}}, \ and\ \bibinfo {author} {\bibfnamefont {B.}~\bibnamefont
  {Carter}},\ }\href {\doibase 10.1046/j.1365-8711.1998.01575.x} {\bibfield
  {journal} {\bibinfo  {journal} {\mnras}\ }\textbf {\bibinfo {volume} {297}},\
  \bibinfo {pages} {1189} (\bibinfo {year} {1998})},\ \Eprint
  {http://arxiv.org/abs/astro-ph/9711042} {astro-ph/9711042} \BibitemShut
  {NoStop}%
\bibitem [{\citenamefont {{Andersson}}\ and\ \citenamefont
  {{Comer}}(2001)}]{andersson_comer2000}%
  \BibitemOpen
  \bibfield  {author} {\bibinfo {author} {\bibfnamefont {N.}~\bibnamefont
  {{Andersson}}}\ and\ \bibinfo {author} {\bibfnamefont {G.~L.}\ \bibnamefont
  {{Comer}}},\ }\href {\doibase 10.1088/0264-9381/18/6/302} {\bibfield
  {journal} {\bibinfo  {journal} {Classical and Quantum Gravity}\ }\textbf
  {\bibinfo {volume} {18}},\ \bibinfo {pages} {969} (\bibinfo {year} {2001})},\
  \Eprint {http://arxiv.org/abs/gr-qc/0009089} {arXiv:gr-qc/0009089 [gr-qc]}
  \BibitemShut {NoStop}%
\bibitem [{\citenamefont {{Gavassino}}\ \emph {et~al.}(2021)\citenamefont
  {{Gavassino}}, \citenamefont {{Antonelli}},\ and\ \citenamefont
  {{Haskell}}}]{GavassinoIordanskii2021}%
  \BibitemOpen
  \bibfield  {author} {\bibinfo {author} {\bibfnamefont {L.}~\bibnamefont
  {{Gavassino}}}, \bibinfo {author} {\bibfnamefont {M.}~\bibnamefont
  {{Antonelli}}}, \ and\ \bibinfo {author} {\bibfnamefont {B.}~\bibnamefont
  {{Haskell}}},\ }\href {\doibase 10.3390/universe7020028} {\bibfield
  {journal} {\bibinfo  {journal} {Universe}\ }\textbf {\bibinfo {volume} {7}},\
  \bibinfo {pages} {28} (\bibinfo {year} {2021})},\ \Eprint
  {http://arxiv.org/abs/2012.10288} {arXiv:2012.10288 [astro-ph.HE]}
  \BibitemShut {NoStop}%
\bibitem [{\citenamefont {{Sourie}}\ \emph {et~al.}(2017)\citenamefont
  {{Sourie}}, \citenamefont {{Chamel}}, \citenamefont {{Novak}},\ and\
  \citenamefont {{Oertel}}}]{sourie_glitch2017}%
  \BibitemOpen
  \bibfield  {author} {\bibinfo {author} {\bibfnamefont {A.}~\bibnamefont
  {{Sourie}}}, \bibinfo {author} {\bibfnamefont {N.}~\bibnamefont {{Chamel}}},
  \bibinfo {author} {\bibfnamefont {J.}~\bibnamefont {{Novak}}}, \ and\
  \bibinfo {author} {\bibfnamefont {M.}~\bibnamefont {{Oertel}}},\ }\href
  {\doibase 10.1093/mnras/stw2613} {\bibfield  {journal} {\bibinfo  {journal}
  {\mnras}\ }\textbf {\bibinfo {volume} {464}},\ \bibinfo {pages} {4641}
  (\bibinfo {year} {2017})}\BibitemShut {NoStop}%
\bibitem [{\citenamefont {{Antonelli}}\ \emph {et~al.}(2018)\citenamefont
  {{Antonelli}}, \citenamefont {{Montoli}},\ and\ \citenamefont
  {{Pizzochero}}}]{antonelli+2018}%
  \BibitemOpen
  \bibfield  {author} {\bibinfo {author} {\bibfnamefont {M.}~\bibnamefont
  {{Antonelli}}}, \bibinfo {author} {\bibfnamefont {A.}~\bibnamefont
  {{Montoli}}}, \ and\ \bibinfo {author} {\bibfnamefont {P.~M.}\ \bibnamefont
  {{Pizzochero}}},\ }\href {\doibase 10.1093/mnras/sty130} {\bibfield
  {journal} {\bibinfo  {journal} {\mnras}\ }\textbf {\bibinfo {volume} {475}},\
  \bibinfo {pages} {5403} (\bibinfo {year} {2018})},\ \Eprint
  {http://arxiv.org/abs/1710.05879} {arXiv:1710.05879 [astro-ph.HE]}
  \BibitemShut {NoStop}%
\bibitem [{\citenamefont {{Gavassino}}\ \emph {et~al.}(2020)\citenamefont
  {{Gavassino}}, \citenamefont {{Antonelli}}, \citenamefont {{Pizzochero}},\
  and\ \citenamefont {{Haskell}}}]{Geo2020}%
  \BibitemOpen
  \bibfield  {author} {\bibinfo {author} {\bibfnamefont {L.}~\bibnamefont
  {{Gavassino}}}, \bibinfo {author} {\bibfnamefont {M.}~\bibnamefont
  {{Antonelli}}}, \bibinfo {author} {\bibfnamefont {P.~M.}\ \bibnamefont
  {{Pizzochero}}}, \ and\ \bibinfo {author} {\bibfnamefont {B.}~\bibnamefont
  {{Haskell}}},\ }\href {\doibase 10.1093/mnras/staa886} {\bibfield  {journal}
  {\bibinfo  {journal} {\mnras}\ }\textbf {\bibinfo {volume} {494}},\ \bibinfo
  {pages} {3562} (\bibinfo {year} {2020})},\ \Eprint
  {http://arxiv.org/abs/2001.08951} {arXiv:2001.08951 [astro-ph.HE]}
  \BibitemShut {NoStop}%
\bibitem [{\citenamefont {Carter}(1989)}]{noto_rel}%
  \BibitemOpen
  \bibfield  {author} {\bibinfo {author} {\bibfnamefont {B.}~\bibnamefont
  {Carter}},\ }\href {\doibase 10.1007/BFb0084028} {\emph {\bibinfo {title}
  {Covariant theory of conductivity in ideal fluid or solid media}}},\ Vol.\
  \bibinfo {volume} {1385}\ (\bibinfo {year} {1989})\ p.~\bibinfo {pages}
  {1}\BibitemShut {NoStop}%
\bibitem [{\citenamefont {{Lopez-Monsalvo}}\ and\ \citenamefont
  {{Andersson}}(2011)}]{Lopez09}%
  \BibitemOpen
  \bibfield  {author} {\bibinfo {author} {\bibfnamefont {C.~S.}\ \bibnamefont
  {{Lopez-Monsalvo}}}\ and\ \bibinfo {author} {\bibfnamefont {N.}~\bibnamefont
  {{Andersson}}},\ }\href {\doibase 10.1098/rspa.2010.0308} {\bibfield
  {journal} {\bibinfo  {journal} {Proceedings of the Royal Society of London
  Series A}\ }\textbf {\bibinfo {volume} {467}},\ \bibinfo {pages} {738}
  (\bibinfo {year} {2011})},\ \Eprint {http://arxiv.org/abs/1006.2978}
  {arXiv:1006.2978 [gr-qc]} \BibitemShut {NoStop}%
\bibitem [{\citenamefont {{Andersson}}\ and\ \citenamefont
  {{Lopez-Monsalvo}}(2011)}]{Lopez11}%
  \BibitemOpen
  \bibfield  {author} {\bibinfo {author} {\bibfnamefont {N.}~\bibnamefont
  {{Andersson}}}\ and\ \bibinfo {author} {\bibfnamefont {C.~S.}\ \bibnamefont
  {{Lopez-Monsalvo}}},\ }\href {\doibase 10.1088/0264-9381/28/19/195023}
  {\bibfield  {journal} {\bibinfo  {journal} {Classical and Quantum Gravity}\
  }\textbf {\bibinfo {volume} {28}},\ \bibinfo {eid} {195023} (\bibinfo {year}
  {2011})},\ \Eprint {http://arxiv.org/abs/1107.0165} {arXiv:1107.0165 [gr-qc]}
  \BibitemShut {NoStop}%
\bibitem [{\citenamefont {Jou}\ \emph {et~al.}(1999)\citenamefont {Jou},
  \citenamefont {Casas-Vazquez},\ and\ \citenamefont {Lebon}}]{Jou_Extended}%
  \BibitemOpen
  \bibfield  {author} {\bibinfo {author} {\bibfnamefont {D.}~\bibnamefont
  {Jou}}, \bibinfo {author} {\bibfnamefont {J.}~\bibnamefont {Casas-Vazquez}},
  \ and\ \bibinfo {author} {\bibfnamefont {G.}~\bibnamefont {Lebon}},\ }\href
  {\doibase 10.1088/0034-4885/51/8/002} {\bibfield  {journal} {\bibinfo
  {journal} {Reports on Progress in Physics}\ }\textbf {\bibinfo {volume}
  {51}},\ \bibinfo {pages} {1105} (\bibinfo {year} {1999})}\BibitemShut
  {NoStop}%
\bibitem [{\citenamefont {Gavassino}\ \emph {et~al.}(2021)\citenamefont
  {Gavassino}, \citenamefont {Antonelli},\ and\ \citenamefont
  {Haskell}}]{BulkGavassino}%
  \BibitemOpen
  \bibfield  {author} {\bibinfo {author} {\bibfnamefont {L.}~\bibnamefont
  {Gavassino}}, \bibinfo {author} {\bibfnamefont {M.}~\bibnamefont
  {Antonelli}}, \ and\ \bibinfo {author} {\bibfnamefont {B.}~\bibnamefont
  {Haskell}},\ }\href {\doibase 10.1088/1361-6382/abe588} {\bibfield  {journal}
  {\bibinfo  {journal} {Classical and Quantum Gravity}\ }\textbf {\bibinfo
  {volume} {38}},\ \bibinfo {pages} {075001} (\bibinfo {year}
  {2021})}\BibitemShut {NoStop}%
\bibitem [{\citenamefont {{Gavassino}}\ and\ \citenamefont
  {{Antonelli}}(2021)}]{GavassinoUEIT2021}%
  \BibitemOpen
  \bibfield  {author} {\bibinfo {author} {\bibfnamefont {L.}~\bibnamefont
  {{Gavassino}}}\ and\ \bibinfo {author} {\bibfnamefont {M.}~\bibnamefont
  {{Antonelli}}},\ }\href {\doibase 10.3389/fspas.2021.686344} {\bibfield
  {journal} {\bibinfo  {journal} {Frontiers in Astronomy and Space Sciences}\
  }\textbf {\bibinfo {volume} {8}},\ \bibinfo {eid} {92} (\bibinfo {year}
  {2021})},\ \Eprint {http://arxiv.org/abs/2105.15184} {arXiv:2105.15184
  [gr-qc]} \BibitemShut {NoStop}%
\bibitem [{\citenamefont {{Camelio}}\ \emph {et~al.}(2022)\citenamefont
  {{Camelio}}, \citenamefont {{Gavassino}}, \citenamefont {{Antonelli}},
  \citenamefont {{Bernuzzi}},\ and\ \citenamefont
  {{Haskell}}}]{CamelioGavassino2022}%
  \BibitemOpen
  \bibfield  {author} {\bibinfo {author} {\bibfnamefont {G.}~\bibnamefont
  {{Camelio}}}, \bibinfo {author} {\bibfnamefont {L.}~\bibnamefont
  {{Gavassino}}}, \bibinfo {author} {\bibfnamefont {M.}~\bibnamefont
  {{Antonelli}}}, \bibinfo {author} {\bibfnamefont {S.}~\bibnamefont
  {{Bernuzzi}}}, \ and\ \bibinfo {author} {\bibfnamefont {B.}~\bibnamefont
  {{Haskell}}},\ }\href@noop {} {\bibfield  {journal} {\bibinfo  {journal}
  {arXiv e-prints}\ ,\ \bibinfo {eid} {arXiv:2204.11809}} (\bibinfo {year}
  {2022})},\ \Eprint {http://arxiv.org/abs/2204.11809} {arXiv:2204.11809
  [gr-qc]} \BibitemShut {NoStop}%
\bibitem [{\citenamefont {Olson}\ and\ \citenamefont
  {Hiscock}(1990)}]{OlsonRegularCarter1990}%
  \BibitemOpen
  \bibfield  {author} {\bibinfo {author} {\bibfnamefont {T.~S.}\ \bibnamefont
  {Olson}}\ and\ \bibinfo {author} {\bibfnamefont {W.~A.}\ \bibnamefont
  {Hiscock}},\ }\href {\doibase 10.1103/PhysRevD.41.3687} {\bibfield  {journal}
  {\bibinfo  {journal} {Phys. Rev. D}\ }\textbf {\bibinfo {volume} {41}},\
  \bibinfo {pages} {3687} (\bibinfo {year} {1990})}\BibitemShut {NoStop}%
\bibitem [{\citenamefont {Priou}(1991)}]{PriouCOMPAR1991}%
  \BibitemOpen
  \bibfield  {author} {\bibinfo {author} {\bibfnamefont {D.}~\bibnamefont
  {Priou}},\ }\href {\doibase 10.1103/PhysRevD.43.1223} {\bibfield  {journal}
  {\bibinfo  {journal} {Phys. Rev. D}\ }\textbf {\bibinfo {volume} {43}},\
  \bibinfo {pages} {1223} (\bibinfo {year} {1991})}\BibitemShut {NoStop}%
\bibitem [{\citenamefont {Israel}\ and\ \citenamefont
  {Stewart}(1979)}]{Israel_Stewart_1979}%
  \BibitemOpen
  \bibfield  {author} {\bibinfo {author} {\bibfnamefont {W.}~\bibnamefont
  {Israel}}\ and\ \bibinfo {author} {\bibfnamefont {J.}~\bibnamefont
  {Stewart}},\ }\href {\doibase https://doi.org/10.1016/0003-4916(79)90130-1}
  {\bibfield  {journal} {\bibinfo  {journal} {Annals of Physics}\ }\textbf
  {\bibinfo {volume} {118}},\ \bibinfo {pages} {341 } (\bibinfo {year}
  {1979})}\BibitemShut {NoStop}%
\bibitem [{\citenamefont {Hiscock}\ and\ \citenamefont
  {Lindblom}(1983)}]{Hishcock1983}%
  \BibitemOpen
  \bibfield  {author} {\bibinfo {author} {\bibfnamefont {W.~A.}\ \bibnamefont
  {Hiscock}}\ and\ \bibinfo {author} {\bibfnamefont {L.}~\bibnamefont
  {Lindblom}},\ }\href {\doibase https://doi.org/10.1016/0003-4916(83)90288-9}
  {\bibfield  {journal} {\bibinfo  {journal} {Annals of Physics}\ }\textbf
  {\bibinfo {volume} {151}},\ \bibinfo {pages} {466 } (\bibinfo {year}
  {1983})}\BibitemShut {NoStop}%
\bibitem [{\citenamefont {Eckart}(1940)}]{Eckart40}%
  \BibitemOpen
  \bibfield  {author} {\bibinfo {author} {\bibfnamefont {C.}~\bibnamefont
  {Eckart}},\ }\href {\doibase 10.1103/PhysRev.58.919} {\bibfield  {journal}
  {\bibinfo  {journal} {Phys. Rev.}\ }\textbf {\bibinfo {volume} {58}},\
  \bibinfo {pages} {919} (\bibinfo {year} {1940})}\BibitemShut {NoStop}%
\bibitem [{\citenamefont {Landau}\ and\ \citenamefont
  {Lifshitz}(2013{\natexlab{a}})}]{landau6}%
  \BibitemOpen
  \bibfield  {author} {\bibinfo {author} {\bibfnamefont {L.}~\bibnamefont
  {Landau}}\ and\ \bibinfo {author} {\bibfnamefont {E.}~\bibnamefont
  {Lifshitz}},\ }\href@noop {} {\emph {\bibinfo {title} {Fluid Mechanics}}},\
  \bibinfo {number} {v. 6}\ (\bibinfo  {publisher} {Elsevier Science},\
  \bibinfo {year} {2013})\BibitemShut {NoStop}%
\bibitem [{\citenamefont {{Prix}}(2004)}]{prix2004}%
  \BibitemOpen
  \bibfield  {author} {\bibinfo {author} {\bibfnamefont {R.}~\bibnamefont
  {{Prix}}},\ }\href {\doibase 10.1103/PhysRevD.69.043001} {\bibfield
  {journal} {\bibinfo  {journal} {\prd}\ }\textbf {\bibinfo {volume} {69}},\
  \bibinfo {eid} {043001} (\bibinfo {year} {2004})},\ \Eprint
  {http://arxiv.org/abs/physics/0209024} {physics/0209024} \BibitemShut
  {NoStop}%
\bibitem [{\citenamefont {{Gavassino}}\ \emph {et~al.}(2022)\citenamefont
  {{Gavassino}}, \citenamefont {{Antonelli}},\ and\ \citenamefont
  {{Haskell}}}]{GavassinoKhalatnikov2021}%
  \BibitemOpen
  \bibfield  {author} {\bibinfo {author} {\bibfnamefont {L.}~\bibnamefont
  {{Gavassino}}}, \bibinfo {author} {\bibfnamefont {M.}~\bibnamefont
  {{Antonelli}}}, \ and\ \bibinfo {author} {\bibfnamefont {B.}~\bibnamefont
  {{Haskell}}},\ }\href {\doibase 10.1103/PhysRevD.105.045011} {\bibfield
  {journal} {\bibinfo  {journal} {\prd}\ }\textbf {\bibinfo {volume} {105}},\
  \bibinfo {eid} {045011} (\bibinfo {year} {2022})},\ \Eprint
  {http://arxiv.org/abs/2110.05546} {arXiv:2110.05546 [gr-qc]} \BibitemShut
  {NoStop}%
\bibitem [{\citenamefont
  {{Gavassino}}(2021{\natexlab{a}})}]{GavassinoGibbs2021}%
  \BibitemOpen
  \bibfield  {author} {\bibinfo {author} {\bibfnamefont {L.}~\bibnamefont
  {{Gavassino}}},\ }\href {\doibase 10.1088/1361-6382/ac2b0e} {\bibfield
  {journal} {\bibinfo  {journal} {Classical and Quantum Gravity}\ }\textbf
  {\bibinfo {volume} {38}},\ \bibinfo {eid} {21LT02} (\bibinfo {year}
  {2021}{\natexlab{a}})},\ \Eprint {http://arxiv.org/abs/2104.09142}
  {arXiv:2104.09142 [gr-qc]} \BibitemShut {NoStop}%
\bibitem [{\citenamefont {Gavassino}\ \emph
  {et~al.}(2020{\natexlab{a}})\citenamefont {Gavassino}, \citenamefont
  {Antonelli},\ and\ \citenamefont {Haskell}}]{GavassinoLyapunov_2020}%
  \BibitemOpen
  \bibfield  {author} {\bibinfo {author} {\bibfnamefont {L.}~\bibnamefont
  {Gavassino}}, \bibinfo {author} {\bibfnamefont {M.}~\bibnamefont
  {Antonelli}}, \ and\ \bibinfo {author} {\bibfnamefont {B.}~\bibnamefont
  {Haskell}},\ }\href {\doibase 10.1103/physrevd.102.043018} {\bibfield
  {journal} {\bibinfo  {journal} {Physical Review D}\ }\textbf {\bibinfo
  {volume} {102}} (\bibinfo {year} {2020}{\natexlab{a}}),\
  10.1103/physrevd.102.043018}\BibitemShut {NoStop}%
\bibitem [{\citenamefont {Gavassino}\ \emph {et~al.}(2022)\citenamefont
  {Gavassino}, \citenamefont {Antonelli},\ and\ \citenamefont
  {Haskell}}]{GavassinoCausality2021}%
  \BibitemOpen
  \bibfield  {author} {\bibinfo {author} {\bibfnamefont {L.}~\bibnamefont
  {Gavassino}}, \bibinfo {author} {\bibfnamefont {M.}~\bibnamefont
  {Antonelli}}, \ and\ \bibinfo {author} {\bibfnamefont {B.}~\bibnamefont
  {Haskell}},\ }\href {\doibase 10.1103/PhysRevLett.128.010606} {\bibfield
  {journal} {\bibinfo  {journal} {Phys. Rev. Lett.}\ }\textbf {\bibinfo
  {volume} {128}},\ \bibinfo {pages} {010606} (\bibinfo {year}
  {2022})}\BibitemShut {NoStop}%
\bibitem [{\citenamefont
  {{Gavassino}}(2021{\natexlab{b}})}]{GavassinoSuperluminal2021}%
  \BibitemOpen
  \bibfield  {author} {\bibinfo {author} {\bibfnamefont {L.}~\bibnamefont
  {{Gavassino}}},\ }\href@noop {} {\bibfield  {journal} {\bibinfo  {journal}
  {arXiv e-prints}\ ,\ \bibinfo {eid} {arXiv:2111.05254}} (\bibinfo {year}
  {2021}{\natexlab{b}})},\ \Eprint {http://arxiv.org/abs/2111.05254}
  {arXiv:2111.05254 [gr-qc]} \BibitemShut {NoStop}%
\bibitem [{\citenamefont {{Bemfica}}\ \emph {et~al.}(2020)\citenamefont
  {{Bemfica}}, \citenamefont {{Disconzi}},\ and\ \citenamefont
  {{Noronha}}}]{BemficaDNDefinitivo2020}%
  \BibitemOpen
  \bibfield  {author} {\bibinfo {author} {\bibfnamefont {F.~S.}\ \bibnamefont
  {{Bemfica}}}, \bibinfo {author} {\bibfnamefont {M.~M.}\ \bibnamefont
  {{Disconzi}}}, \ and\ \bibinfo {author} {\bibfnamefont {J.}~\bibnamefont
  {{Noronha}}},\ }\href@noop {} {\bibfield  {journal} {\bibinfo  {journal}
  {arXiv e-prints}\ ,\ \bibinfo {eid} {arXiv:2009.11388}} (\bibinfo {year}
  {2020})},\ \Eprint {http://arxiv.org/abs/2009.11388} {arXiv:2009.11388
  [gr-qc]} \BibitemShut {NoStop}%
\bibitem [{\citenamefont {{Lebedev}}\ and\ \citenamefont
  {{Khalatnikov}}(1982)}]{lebedev1982}%
  \BibitemOpen
  \bibfield  {author} {\bibinfo {author} {\bibfnamefont {V.~V.}\ \bibnamefont
  {{Lebedev}}}\ and\ \bibinfo {author} {\bibfnamefont {I.~M.}\ \bibnamefont
  {{Khalatnikov}}},\ }\href@noop {} {\bibfield  {journal} {\bibinfo  {journal}
  {Zhurnal Eksperimentalnoi i Teoreticheskoi Fiziki}\ }\textbf {\bibinfo
  {volume} {83}},\ \bibinfo {pages} {1601} (\bibinfo {year}
  {1982})}\BibitemShut {NoStop}%
\bibitem [{\citenamefont {{Misner}}\ \emph {et~al.}(1973)\citenamefont
  {{Misner}}, \citenamefont {{Thorne}},\ and\ \citenamefont
  {{Wheeler}}}]{MTW_book}%
  \BibitemOpen
  \bibfield  {author} {\bibinfo {author} {\bibfnamefont {C.~W.}\ \bibnamefont
  {{Misner}}}, \bibinfo {author} {\bibfnamefont {K.~S.}\ \bibnamefont
  {{Thorne}}}, \ and\ \bibinfo {author} {\bibfnamefont {J.~A.}\ \bibnamefont
  {{Wheeler}}},\ }\href@noop {} {\emph {\bibinfo {title} {San Francisco:
  W.H.~Freeman and Co., 1973}}}\ (\bibinfo {year} {1973})\BibitemShut {NoStop}%
\bibitem [{\citenamefont {Gavassino}(2020)}]{GavassinoTermometri}%
  \BibitemOpen
  \bibfield  {author} {\bibinfo {author} {\bibfnamefont {L.}~\bibnamefont
  {Gavassino}},\ }\href {\doibase 10.1007/s10701-020-00393-x} {\bibfield
  {journal} {\bibinfo  {journal} {Found. Phys.}\ }\textbf {\bibinfo {volume}
  {50}},\ \bibinfo {pages} {1554} (\bibinfo {year} {2020})},\ \Eprint
  {http://arxiv.org/abs/2005.06396} {arXiv:2005.06396 [gr-qc]} \BibitemShut
  {NoStop}%
\bibitem [{\citenamefont {Hawking}\ and\ \citenamefont
  {Ellis}(2011)}]{Hawking1973}%
  \BibitemOpen
  \bibfield  {author} {\bibinfo {author} {\bibfnamefont {S.~W.}\ \bibnamefont
  {Hawking}}\ and\ \bibinfo {author} {\bibfnamefont {G.~F.~R.}\ \bibnamefont
  {Ellis}},\ }\href {\doibase 10.1017/CBO9780511524646} {\emph {\bibinfo
  {title} {{The Large Scale Structure of Space-Time}}}},\ Cambridge Monographs
  on Mathematical Physics\ (\bibinfo  {publisher} {Cambridge University
  Press},\ \bibinfo {year} {2011})\BibitemShut {NoStop}%
\bibitem [{\citenamefont {{Becattini}}(2016)}]{Becattini2016}%
  \BibitemOpen
  \bibfield  {author} {\bibinfo {author} {\bibfnamefont {F.}~\bibnamefont
  {{Becattini}}},\ }\href {\doibase 10.5506/APhysPolB.47.1819} {\bibfield
  {journal} {\bibinfo  {journal} {Acta Physica Polonica B}\ }\textbf {\bibinfo
  {volume} {47}},\ \bibinfo {pages} {1819} (\bibinfo {year} {2016})},\ \Eprint
  {http://arxiv.org/abs/1606.06605} {arXiv:1606.06605 [gr-qc]} \BibitemShut
  {NoStop}%
\bibitem [{\citenamefont {{Andreev}}\ and\ \citenamefont
  {{Melnikovsky}}(2004)}]{AndreevMelnikovski2004}%
  \BibitemOpen
  \bibfield  {author} {\bibinfo {author} {\bibfnamefont {A.~F.}\ \bibnamefont
  {{Andreev}}}\ and\ \bibinfo {author} {\bibfnamefont {L.~A.}\ \bibnamefont
  {{Melnikovsky}}},\ }\href {\doibase 10.1023/B:JOLT.0000029505.92429.f6}
  {\bibfield  {journal} {\bibinfo  {journal} {Journal of Low Temperature
  Physics}\ }\textbf {\bibinfo {volume} {135}},\ \bibinfo {pages} {411}
  (\bibinfo {year} {2004})},\ \Eprint {http://arxiv.org/abs/cond-mat/0405111}
  {arXiv:cond-mat/0405111 [cond-mat.soft]} \BibitemShut {NoStop}%
\bibitem [{\citenamefont {Gibbons}\ and\ \citenamefont
  {Hawking}(1977)}]{GibbonsHawking1977}%
  \BibitemOpen
  \bibfield  {author} {\bibinfo {author} {\bibfnamefont {G.~W.}\ \bibnamefont
  {Gibbons}}\ and\ \bibinfo {author} {\bibfnamefont {S.~W.}\ \bibnamefont
  {Hawking}},\ }\href {\doibase 10.1103/PhysRevD.15.2752} {\bibfield  {journal}
  {\bibinfo  {journal} {Phys. Rev. D}\ }\textbf {\bibinfo {volume} {15}},\
  \bibinfo {pages} {2752} (\bibinfo {year} {1977})}\BibitemShut {NoStop}%
\bibitem [{\citenamefont {Gavassino}\ \emph
  {et~al.}(2020{\natexlab{b}})\citenamefont {Gavassino}, \citenamefont
  {Antonelli},\ and\ \citenamefont {Haskell}}]{GavassinoRadiazione}%
  \BibitemOpen
  \bibfield  {author} {\bibinfo {author} {\bibfnamefont {L.}~\bibnamefont
  {Gavassino}}, \bibinfo {author} {\bibfnamefont {M.}~\bibnamefont
  {Antonelli}}, \ and\ \bibinfo {author} {\bibfnamefont {B.}~\bibnamefont
  {Haskell}},\ }\href {\doibase 10.3390/sym12091543} {\bibfield  {journal}
  {\bibinfo  {journal} {Symmetry}\ }\textbf {\bibinfo {volume} {12}},\ \bibinfo
  {pages} {1543} (\bibinfo {year} {2020}{\natexlab{b}})}\BibitemShut {NoStop}%
\bibitem [{\citenamefont {{Florkowski}}\ \emph {et~al.}(2018)\citenamefont
  {{Florkowski}}, \citenamefont {{Heller}},\ and\ \citenamefont
  {{Spali{\'n}ski}}}]{FlorkowskiReview2018}%
  \BibitemOpen
  \bibfield  {author} {\bibinfo {author} {\bibfnamefont {W.}~\bibnamefont
  {{Florkowski}}}, \bibinfo {author} {\bibfnamefont {M.~P.}\ \bibnamefont
  {{Heller}}}, \ and\ \bibinfo {author} {\bibfnamefont {M.}~\bibnamefont
  {{Spali{\'n}ski}}},\ }\href {\doibase 10.1088/1361-6633/aaa091} {\bibfield
  {journal} {\bibinfo  {journal} {Reports on Progress in Physics}\ }\textbf
  {\bibinfo {volume} {81}},\ \bibinfo {eid} {046001} (\bibinfo {year}
  {2018})},\ \Eprint {http://arxiv.org/abs/1707.02282} {arXiv:1707.02282
  [hep-ph]} \BibitemShut {NoStop}%
\bibitem [{\citenamefont {Carter}\ and\ \citenamefont
  {Khalatnikov}(1992)}]{carter92}%
  \BibitemOpen
  \bibfield  {author} {\bibinfo {author} {\bibfnamefont {B.}~\bibnamefont
  {Carter}}\ and\ \bibinfo {author} {\bibfnamefont {I.}~\bibnamefont
  {Khalatnikov}},\ }\href {\doibase
  https://doi.org/10.1016/0003-4916(92)90348-P} {\bibfield  {journal} {\bibinfo
   {journal} {Annals of Physics}\ }\textbf {\bibinfo {volume} {219}},\ \bibinfo
  {pages} {243 } (\bibinfo {year} {1992})}\BibitemShut {NoStop}%
\bibitem [{\citenamefont {Landau}\ \emph {et~al.}(1980)\citenamefont {Landau},
  \citenamefont {Lifshitz},\ and\ \citenamefont {Pitaevskij}}]{landau9}%
  \BibitemOpen
  \bibfield  {author} {\bibinfo {author} {\bibfnamefont {L.}~\bibnamefont
  {Landau}}, \bibinfo {author} {\bibfnamefont {E.}~\bibnamefont {Lifshitz}}, \
  and\ \bibinfo {author} {\bibfnamefont {L.}~\bibnamefont {Pitaevskij}},\
  }\href {https://books.google.pl/books?id=dEVtKQEACAAJ} {\emph {\bibinfo
  {title} {Statistical Physics: Part 2 : Theory of Condensed State}}},\ Landau
  and Lifshitz Course of theoretical physics\ (\bibinfo  {publisher} {Oxford},\
  \bibinfo {year} {1980})\BibitemShut {NoStop}%
\bibitem [{\citenamefont {{Chamel}}\ and\ \citenamefont
  {{Haensel}}(2006)}]{chamelhaensel2006}%
  \BibitemOpen
  \bibfield  {author} {\bibinfo {author} {\bibfnamefont {N.}~\bibnamefont
  {{Chamel}}}\ and\ \bibinfo {author} {\bibfnamefont {P.}~\bibnamefont
  {{Haensel}}},\ }\href {\doibase 10.1103/PhysRevC.73.045802} {\bibfield
  {journal} {\bibinfo  {journal} {\prc}\ }\textbf {\bibinfo {volume} {73}},\
  \bibinfo {eid} {045802} (\bibinfo {year} {2006})},\ \Eprint
  {http://arxiv.org/abs/nucl-th/0603018} {nucl-th/0603018} \BibitemShut
  {NoStop}%
\bibitem [{\citenamefont {{Carter}}\ \emph {et~al.}(2006)\citenamefont
  {{Carter}}, \citenamefont {{Chamel}},\ and\ \citenamefont
  {{Haensel}}}]{carter_macro_2006}%
  \BibitemOpen
  \bibfield  {author} {\bibinfo {author} {\bibfnamefont {B.}~\bibnamefont
  {{Carter}}}, \bibinfo {author} {\bibfnamefont {N.}~\bibnamefont {{Chamel}}},
  \ and\ \bibinfo {author} {\bibfnamefont {P.}~\bibnamefont {{Haensel}}},\
  }\href {\doibase 10.1142/S0218271806008504} {\bibfield  {journal} {\bibinfo
  {journal} {International Journal of Modern Physics D}\ }\textbf {\bibinfo
  {volume} {15}},\ \bibinfo {pages} {777} (\bibinfo {year} {2006})},\ \Eprint
  {http://arxiv.org/abs/astro-ph/0408083} {arXiv:astro-ph/0408083 [astro-ph]}
  \BibitemShut {NoStop}%
\bibitem [{\citenamefont {Rau}\ and\ \citenamefont
  {Wasserman}(2020)}]{RauWasserman2020}%
  \BibitemOpen
  \bibfield  {author} {\bibinfo {author} {\bibfnamefont {P.~B.}\ \bibnamefont
  {Rau}}\ and\ \bibinfo {author} {\bibfnamefont {I.}~\bibnamefont
  {Wasserman}},\ }\href {\doibase 10.1103/PhysRevD.102.063011} {\bibfield
  {journal} {\bibinfo  {journal} {Phys. Rev. D}\ }\textbf {\bibinfo {volume}
  {102}},\ \bibinfo {pages} {063011} (\bibinfo {year} {2020})}\BibitemShut
  {NoStop}%
\bibitem [{\citenamefont {Landau}\ and\ \citenamefont
  {Lifshitz}(2013{\natexlab{b}})}]{landau5}%
  \BibitemOpen
  \bibfield  {author} {\bibinfo {author} {\bibfnamefont {L.}~\bibnamefont
  {Landau}}\ and\ \bibinfo {author} {\bibfnamefont {E.}~\bibnamefont
  {Lifshitz}},\ }\href {https://books.google.pl/books?id=VzgJN-XPTRsC} {\emph
  {\bibinfo {title} {Statistical Physics}}},\ \bibinfo {number} {v. 5}\
  (\bibinfo  {publisher} {Elsevier Science},\ \bibinfo {year}
  {2013})\BibitemShut {NoStop}%
\bibitem [{\citenamefont {Wald}(1984)}]{Wald}%
  \BibitemOpen
  \bibfield  {author} {\bibinfo {author} {\bibfnamefont {R.~M.}\ \bibnamefont
  {Wald}},\ }\href {https://cds.cern.ch/record/106274} {\emph {\bibinfo {title}
  {{General relativity}}}}\ (\bibinfo  {publisher} {Chicago Univ. Press},\
  \bibinfo {address} {Chicago, IL},\ \bibinfo {year} {1984})\BibitemShut
  {NoStop}%
\bibitem [{\citenamefont {{Bemfica}}\ \emph {et~al.}(2018)\citenamefont
  {{Bemfica}}, \citenamefont {{Disconzi}},\ and\ \citenamefont
  {{Noronha}}}]{BemficaCausality2018}%
  \BibitemOpen
  \bibfield  {author} {\bibinfo {author} {\bibfnamefont {F.~S.}\ \bibnamefont
  {{Bemfica}}}, \bibinfo {author} {\bibfnamefont {M.~M.}\ \bibnamefont
  {{Disconzi}}}, \ and\ \bibinfo {author} {\bibfnamefont {J.}~\bibnamefont
  {{Noronha}}},\ }\href {\doibase 10.1103/PhysRevD.98.104064} {\bibfield
  {journal} {\bibinfo  {journal} {\prd}\ }\textbf {\bibinfo {volume} {98}},\
  \bibinfo {eid} {104064} (\bibinfo {year} {2018})},\ \Eprint
  {http://arxiv.org/abs/1708.06255} {arXiv:1708.06255 [gr-qc]} \BibitemShut
  {NoStop}%
\bibitem [{\citenamefont {Aharonov}\ \emph {et~al.}(1969)\citenamefont
  {Aharonov}, \citenamefont {Komar},\ and\ \citenamefont
  {Susskind}}]{Susskind1969}%
  \BibitemOpen
  \bibfield  {author} {\bibinfo {author} {\bibfnamefont {Y.}~\bibnamefont
  {Aharonov}}, \bibinfo {author} {\bibfnamefont {A.}~\bibnamefont {Komar}}, \
  and\ \bibinfo {author} {\bibfnamefont {L.}~\bibnamefont {Susskind}},\ }\href
  {\doibase 10.1103/PhysRev.182.1400} {\bibfield  {journal} {\bibinfo
  {journal} {Phys. Rev.}\ }\textbf {\bibinfo {volume} {182}},\ \bibinfo {pages}
  {1400} (\bibinfo {year} {1969})}\BibitemShut {NoStop}%
\bibitem [{\citenamefont {{Krotscheck}}\ and\ \citenamefont
  {{Kundt}}(1978)}]{Krotscheck1978}%
  \BibitemOpen
  \bibfield  {author} {\bibinfo {author} {\bibfnamefont {E.}~\bibnamefont
  {{Krotscheck}}}\ and\ \bibinfo {author} {\bibfnamefont {W.}~\bibnamefont
  {{Kundt}}},\ }\href {\doibase 10.1007/BF01609447} {\bibfield  {journal}
  {\bibinfo  {journal} {Communications in Mathematical Physics}\ }\textbf
  {\bibinfo {volume} {60}},\ \bibinfo {pages} {171} (\bibinfo {year}
  {1978})}\BibitemShut {NoStop}%
\bibitem [{\citenamefont {{Fox}}\ \emph {et~al.}(1970)\citenamefont {{Fox}},
  \citenamefont {{Kuper}},\ and\ \citenamefont {{Lipson}}}]{FoxKuper1970}%
  \BibitemOpen
  \bibfield  {author} {\bibinfo {author} {\bibfnamefont {R.}~\bibnamefont
  {{Fox}}}, \bibinfo {author} {\bibfnamefont {C.~G.}\ \bibnamefont {{Kuper}}},
  \ and\ \bibinfo {author} {\bibfnamefont {S.~G.}\ \bibnamefont {{Lipson}}},\
  }\href {\doibase 10.1098/rspa.1970.0093} {\bibfield  {journal} {\bibinfo
  {journal} {Proceedings of the Royal Society of London Series A}\ }\textbf
  {\bibinfo {volume} {316}},\ \bibinfo {pages} {515} (\bibinfo {year}
  {1970})}\BibitemShut {NoStop}%
\bibitem [{\citenamefont {Rauch}(1991)}]{Rauch_book}%
  \BibitemOpen
  \bibfield  {author} {\bibinfo {author} {\bibfnamefont {J.}~\bibnamefont
  {Rauch}},\ }\href {https://doi.org/10.1007/978-1-4612-0953-9} {\emph
  {\bibinfo {title} {{Partial Differential Equations}}}},\ Graduate Texts in
  Mathematics\ (\bibinfo  {publisher} {Springer},\ \bibinfo {address} {New
  York, NY},\ \bibinfo {year} {1991})\BibitemShut {NoStop}%
\bibitem [{\citenamefont {{Pu}}\ \emph {et~al.}(2010)\citenamefont {{Pu}},
  \citenamefont {{Koide}},\ and\ \citenamefont {{Rischke}}}]{Pu2010}%
  \BibitemOpen
  \bibfield  {author} {\bibinfo {author} {\bibfnamefont {S.}~\bibnamefont
  {{Pu}}}, \bibinfo {author} {\bibfnamefont {T.}~\bibnamefont {{Koide}}}, \
  and\ \bibinfo {author} {\bibfnamefont {D.~H.}\ \bibnamefont {{Rischke}}},\
  }\href {\doibase 10.1103/PhysRevD.81.114039} {\bibfield  {journal} {\bibinfo
  {journal} {\prd}\ }\textbf {\bibinfo {volume} {81}},\ \bibinfo {eid} {114039}
  (\bibinfo {year} {2010})},\ \Eprint {http://arxiv.org/abs/0907.3906}
  {arXiv:0907.3906 [hep-ph]} \BibitemShut {NoStop}%
\bibitem [{\citenamefont {Stueckelberg}(1962)}]{Stuekelberg1962}%
  \BibitemOpen
  \bibfield  {author} {\bibinfo {author} {\bibfnamefont {E.}~\bibnamefont
  {Stueckelberg}},\ }\href@noop {} {\bibfield  {journal} {\bibinfo  {journal}
  {Helvetica Physica Acta}\ }\textbf {\bibinfo {volume} {35}} (\bibinfo {year}
  {1962})}\BibitemShut {NoStop}%
\bibitem [{\citenamefont {Israel}(2009)}]{Israel_2009_inbook}%
  \BibitemOpen
  \bibfield  {author} {\bibinfo {author} {\bibfnamefont {W.}~\bibnamefont
  {Israel}},\ }\enquote {\bibinfo {title} {Relativistic thermodynamics},}\ in\
  \href {\doibase 10.1007/978-3-7643-8878-2_8} {\emph {\bibinfo {booktitle}
  {E.C.G. Stueckelberg, An Unconventional Figure of Twentieth Century Physics:
  Selected Scientific Papers with Commentaries}}},\ \bibinfo {editor} {edited
  by\ \bibinfo {editor} {\bibfnamefont {J.}~\bibnamefont {Lacki}}, \bibinfo
  {editor} {\bibfnamefont {H.}~\bibnamefont {Ruegg}}, \ and\ \bibinfo {editor}
  {\bibfnamefont {G.}~\bibnamefont {Wanders}}}\ (\bibinfo  {publisher}
  {Birkh{\"a}user Basel},\ \bibinfo {address} {Basel},\ \bibinfo {year}
  {2009})\ pp.\ \bibinfo {pages} {101--113}\BibitemShut {NoStop}%
\bibitem [{\citenamefont {Kondepudi}\ and\ \citenamefont
  {Prigogine}(2014)}]{PrigoginebookModernThermodynamics2014}%
  \BibitemOpen
  \bibfield  {author} {\bibinfo {author} {\bibfnamefont {D.}~\bibnamefont
  {Kondepudi}}\ and\ \bibinfo {author} {\bibfnamefont {I.}~\bibnamefont
  {Prigogine}},\ }\href@noop {} {\emph {\bibinfo {title} {Modern
  Thermodynamics}}}\ (\bibinfo  {publisher} {John Wiley and Sons, Ltd},\
  \bibinfo {year} {2014})\BibitemShut {NoStop}%
\bibitem [{\citenamefont {{Rezzolla}}\ and\ \citenamefont
  {{Zanotti}}(2013)}]{rezzolla_book}%
  \BibitemOpen
  \bibfield  {author} {\bibinfo {author} {\bibfnamefont {L.}~\bibnamefont
  {{Rezzolla}}}\ and\ \bibinfo {author} {\bibfnamefont {O.}~\bibnamefont
  {{Zanotti}}},\ }\href@noop {} {\emph {\bibinfo {title} {Relativistic
  Hydrodynamics, by L.~Rezzolla and O.~Zanotti.~Oxford University Press,
  2013.~ISBN-10: 0198528906; ISBN-13: 978-0198528906}}}\ (\bibinfo {year}
  {2013})\BibitemShut {NoStop}%
\bibitem [{\citenamefont {Callen}(1985)}]{Callen_book}%
  \BibitemOpen
  \bibfield  {author} {\bibinfo {author} {\bibfnamefont {H.~B.}\ \bibnamefont
  {Callen}},\ }\href {https://cds.cern.ch/record/450289} {\emph {\bibinfo
  {title} {{Thermodynamics and an introduction to thermostatistics; 2nd
  ed.}}}}\ (\bibinfo  {publisher} {Wiley},\ \bibinfo {address} {New York, NY},\
  \bibinfo {year} {1985})\BibitemShut {NoStop}%
\bibitem [{\citenamefont {{Son}}(2001)}]{Son2001}%
  \BibitemOpen
  \bibfield  {author} {\bibinfo {author} {\bibfnamefont {D.~T.}\ \bibnamefont
  {{Son}}},\ }\href {\doibase 10.1142/S0217751X01009545} {\bibfield  {journal}
  {\bibinfo  {journal} {International Journal of Modern Physics A}\ }\textbf
  {\bibinfo {volume} {16}},\ \bibinfo {pages} {1284} (\bibinfo {year}
  {2001})},\ \Eprint {http://arxiv.org/abs/hep-ph/0011246} {hep-ph/0011246}
  \BibitemShut {NoStop}%
\bibitem [{\citenamefont {{Gusakov}}(2007)}]{Gusakov2007}%
  \BibitemOpen
  \bibfield  {author} {\bibinfo {author} {\bibfnamefont {M.~E.}\ \bibnamefont
  {{Gusakov}}},\ }\href {\doibase 10.1103/PhysRevD.76.083001} {\bibfield
  {journal} {\bibinfo  {journal} {\prd}\ }\textbf {\bibinfo {volume} {76}},\
  \bibinfo {eid} {083001} (\bibinfo {year} {2007})},\ \Eprint
  {http://arxiv.org/abs/0704.1071} {arXiv:0704.1071} \BibitemShut {NoStop}%
\end{thebibliography}%

\label{lastpage}

\end{document}